\documentclass[aps,prd,preprint,tightenlines,groupedaddress,
   showpacs]{revtex4}
\usepackage{epsf,epsfig,graphics,graphicx}
\begin{document}
%\preprint{MMC\#3}
\title{Off-equilibrium dynamics of the primordial perturbations
   in the inflationary universe: the $O(N)$ model}
\author{Wolung Lee}
\email{leewl@phys.sinica.edu.tw}
\author{Yeo-Yie Charng}
\email{charng@phys.sinica.edu.tw}
 \affiliation{Institute of
Physics, Academia Sinica, Taipei, Taiwan 115, R. O. C.}
\author{Da-Shin Lee}
\email{dslee@mail.ndhu.edu.tw} \affiliation{Department of Physics,
National Dong-Hwa University, Hua-lien, Taiwan 974, R. O. C.}
\author{Li-Zhi Fang}
\email{fanglz@physics.arizona.edu} \affiliation{Department of
Physics, University of Arizona,
         Tucson, AZ 85721, U. S. A.}

\date{\today}

\begin{abstract}

Using the $O(N)$ model as an example, we investigate the
self-interaction effects of inflaton on the dynamics of the
primordial perturbations. When taking interactions into account,
it is essential to employ a self-consistent off-equilibrium
formalism to study the evolution of the inflationary background
field  and its fluctuations with the back-reaction effects. Within
the Hartree factorization scheme, we show that the $O(N)$ model
has at least two observable remains left behind the
off-equilibrium processes: the running spectral index of
primordial density perturbations and the correlations between
perturbation modes in phase space. We find that the running of the
spectral index is fully determined by the rate of the energy
transfer from the inflationary background field to its
fluctuations via particle creation processes as well as the
dynamics of the  background field itself. Furthermore, the
amplitude of the field fluctuations turns out to be
scale-dependent due to the off-equilibrium evolution. As a
consequence, the scale-dependence of fluctuations yields a
correlation between the phase space modes of energy density
perturbations, while the one-point function of the fluctuations in
each Hartree mode is still Gaussian. More importantly, the
mode-mode correlation of the primordial perturbations depends upon
the dynamics of the self-interaction {\it as well as} the initial
conditions of the inflation. Hence, we propose that the running
spectral index and the correlation between phase-space modes would
be two observable fossils to probe the epoch of inflation, even
beyond.

\end{abstract}

 \maketitle

\section{Introduction}

In the inflationary scenario, the primordial perturbations of the
universe originate from vacuum fluctuations of the scalar field(s),
the inflaton $\phi$, driving the inflation. If the dynamics of the
fluctuations is approximated by single massless free field during
the inflationary epoch, the power spectrum of curvature perturbations
for a Fourier mode $k$ can be obtained as
%eq1
\begin{equation}
{\cal P}_{\cal R}(k) =\left [\left (\frac{H}{\dot{\phi}}\right )
   \left (\frac{H}{2\pi}\right)\right]^2_{k=aH}\ , \label{ps}
\end{equation}
where the inflationary scale factor $a=\exp{(Ht)}$, and the
Hubble parameter $H = \sqrt{8\pi G V/3}$ are determined by the
potential of the inflaton field $V(\phi)$. The first factor $H/\dot{\phi}$
on the right hand side of Eq. (\ref{ps}) comes from the evolution of the
background inflaton field $\phi$, i.e. the expectation value of the
quantum scalar field. Meanwhile, the second factor $H/2\pi$ is specified
by the variance of the classical $\phi$ field fluctuations,
$\delta \phi_{k}$, at a few Hubble times after horizon-crossing.
With slow-roll conditions, the derivative of the spectral index,
$dn (k)/d\ln k$, turns out to be very small, or even negligible such
that $n(k)=(d\ln{\cal P}_{\cal R}(k)/d\ln k)+1\simeq 1$ is roughly
$k$-independent~\cite{lid}.

However the $k$-dependence, or the running of the spectral index,
revealed by the recently released Wilkinson Microwave Anisotropy
Probe (WMAP) data can be as large as $dn(k)/d\ln k=(d/d\ln k)^2
\ln{\cal P}_{\cal R}(k)\simeq -0.055{\ \rm to \ }-0.077$~\cite{pei}.
Hence, the origin of the primordial perturbations cannot be solely
accommodated with the quantum fluctuations of single free field.
In a slow-roll inflation, one has
$d/d\ln k =(1/H)d/dt = -(1/8\pi G)(V'(\phi)/V(\phi))(d/d\phi)$, and
therefore the derivative of the power spectrum
$(d/d\ln k)^n\ln{\cal P}_{\cal R}(k)$ will no longer be negligible
for the second order ($n=2$) as long as the third or higher
derivatives of the inflaton potential becomes substantial.
To fit in with the running of the spectral index, models beyond
single scalar field with quadratic potential $V(\phi)$ have
been proposed accordingly~\cite{kaw}. In this context, the running
index of the perturbation power spectrum is considered as an
essential feature due to the interactions or the self-interactions
of the inflaton(s).

It should be pointed out here that in deriving the derivative of
the power spectrum (\ref{ps}), the tree level of the effective
potential $V(\phi)$ is implicitly involved. However, when quantum
fluctuations arising from the interactions of the inflaton become
important, this approach may be problematic. In particular, the
one-loop effective potential will turn complex within the region
where the background field $\phi$ is constrained by
$V''(\phi)<0$~\cite{do}. The imaginary part of the effective
potential would inevitably lead to a dynamically unstable
state~\cite{we}. This so-called ``spinodal instability'' will
allow long wavelength fluctuations to grow
nonperturbatively~\cite{huang}. As such, the term $H/2\pi$ in Eq.
(1) is no longer valid when the inflaton suffers from spinodal
instabilities. Therefore, the primordial perturbations must be
dealt consistently with a different method to account for the
amplification of vacuum fluctuations in the presence of spinodal
instabilities.

This motivates us to study the self-interaction effects of the
inflaton on the dynamics of primordial perturbations within a
context of the self-consistent off-equilibrium formalism. We will
use the $O(N)$ model of inflation as an example. When $N=1$, it
reduces to the popular $\lambda\phi^4$ inflationary model which is
disfavored by the WMAP. However, the $O(N)$ model with
spontaneously broken symmetry is still a viable model in quantum
field theory and off-equilibrium statistical mechanics~\cite{sam}.
It has been extensively used in modeling the quantum
off-equilibrium processes in the early universe, as well as the
chiral phase transition in relativistic heavy ion collision, {\it
etc}~\cite{baa}. In particular, the dynamics of quantum
fluctuations of the $O(N)$ model in the large $N$ limit has been
developed with the Hartree-type factorization~\cite{boyn1,boyn2},
which can be generalized for the finite $N$ situation. In this
paper, we will focus on searching for the possible observable
imprints caused by the off-equilibrium evolution of inflationary
primordial perturbations.

The paper is organized as follows. In Section 2, we briefly
describe the $O(N)$ model of inflation, and introduce the
self-consistent, nonperturbative and renormalized solutions to
the nonlinear evolution of the background inflaton field as well
as its quantum fluctuations within the Hartree approximation.
Section 3 analyzes the dynamics of quantum fluctuations undergoing
a quantum-to-classical transition, and the statistical properties
of the corresponding classical primordial perturbations. The
spectral index along with the running of the perturbation power
spectrum will be calculated in Section 4. Section 5 then discusses
the correlation of primordial perturbations in phase space and its
detection. Finally, we summarize our findings and give conclusions
in Section 6.

\section{THE $O(N)$ INFLATION}

\subsection{The model and the Hartree factorization}

Consider the dynamics of an inflation driven by a field ${\bf\Phi}$
of the $O(N)$ vector model with spontaneous symmetry breaking. The
action is defined by
%eq2
\begin{equation}
{\mathcal S}=\int d^4x {\mathcal L}= \int d^4x\sqrt{-g}\left [
\frac{1}{2}g^{\mu\nu}
     \partial_{\mu} {\bf \Phi} \cdot \partial_{\nu}{\bf \Phi} -
        V({\bf \Phi}\cdot{\bf \Phi})\right ],
\end{equation}
where $V({\bf \Phi}\cdot{\bf \Phi})$ is a self-interaction
potential given by
%eq3
\begin{equation}
V({\bf \Phi}\cdot{\bf \Phi}) = \frac{\lambda}{8N} \left ({\bf
\Phi}\cdot{\bf \Phi} - \frac{2Nm^2}{\lambda}  \right )^2 \, .
\label{pot}
\end{equation}
 As an inflaton, the $N$ components of the field generally are
represented as ${\bf \Phi} = (\sigma, \vec{\pi})$, where $\vec{
\pi} $ represent  $N-1$ scalar fields. The cosmic inflation is
characterized by the state in which the component $\sigma$ has a
spatially homogeneous expectation, i.e. $\sigma$ can be decomposed
into a background plus fluctuations around the background as
%eq4
\begin{equation}
 \sigma({\bf x},t) =\sqrt{N}\phi(t) + \chi({\bf x},t) \,
\end{equation}
with the expectation value  $\langle\sigma({\bf x},t)\rangle=
\sqrt{N}\phi(t)$ and thus,  $\langle \chi ({\bf x},t) \rangle =0$.
During the inflationary epoch, the background space-time can be
described by a spatially flat Friedmann-Robertson-Walker metric,
%eq5
\begin{equation}
ds^2=  g_{\mu\nu} dx^{\mu} dx^{\nu}= dt^2 -
a^2(t)\delta_{ij}dx^idx^j,
\end{equation}
where the scale factor $a=\exp({Ht})$ with the expansion rate
$H=\sqrt{8\pi G\rho/3}$ determined by the mean of energy density
of the inflaton field as
%eq6
\begin{equation}
 H^2 (t) =  \frac{8 \pi N }{3 M^2_{\rm Pl}} \,
\left[ \frac{1}{2} \dot \phi^2
 (t)+ \frac{\lambda}{8} \left(
 \phi^2 (t) -\frac{2 m^2}{\lambda}
 \right)^2 \right], \label{hubbleeq}
\end{equation}
where $ M_{\rm Pl}$ is the Planck mass.  We have assumed that the
effects of quantum fluctuations  on the dynamics of the Hubble
parameter can be ignored  as we will justify this assumption
later.

During the inflation, the expectation value of the scalar field
$\phi(t)$ undergoes an off-equilibrium evolution from the initial
state $\phi(t)\sim 0$ where $V''(\phi) <0$. Accordingly, the mass
square of the long wavelength fluctuation modes will be negative,
which leads to the nonperturbative growth of fluctuations.
Therefore, a nonperturbative framework is necessary for taking
account of the growth of fluctuations, especially in computing the
power spectrum of perturbations later. We will employ the method
of the Hartree factorization, which approximates the potential
$V(\phi)$ with an effective quadratic potential while keeping $N$
finite~\cite{boyn2,baa}. The Hartree-factorized Lagrangian is
%eq7
\begin{equation}
{\mathcal L}(t)= \int d^3 x a^3 (t) \left[ \frac{1}{2}
(\partial_{\mu}\chi )^2 + \frac{1}{2} ( \partial_{\mu}
\vec{\pi})^2 - \frac{1}{2} M^2_{\chi} (t) \chi^2 - \frac{1}{2}
M^2_{\pi }(t) \vec{\pi}^2 - \chi V'(t)\right], \label{hartreeH}
\end{equation}
where
%eq8
\begin{equation}
 V'(t)= \sqrt{N} \left[ \ddot{ \phi} (t) + 3H \dot{\phi} (t) +
     (M_{\chi}^2-\lambda\phi^2 (t) )\phi (t)\right].
\label{potentialchi}
\end{equation}
The time dependent effective masses $M_{\chi}(t) $ and $M_{\pi}
(t)$ are obtained as
%eq9
\begin{eqnarray}
M^2_{\chi}(t) &=& -m^2+\frac{3\lambda}{2}\phi^2 (t)+
\frac{3\lambda}{2N}\langle\chi^2\rangle (t) +
\frac{\lambda}{2}\left(1-\frac{1}{N}\right)\langle \psi^2\rangle
(t) \, ,
\nonumber \\
 M^2_{\pi} (t) &=& -m^2+\frac{\lambda}{2}\phi^2(t) +
\frac{\lambda}{2N}\langle\chi^2\rangle (t) +
\frac{\lambda}{2}\left(1+\frac{1}{N}\right)\langle \psi^2\rangle
(t)  \, , \label{selfconmass}
\end{eqnarray}
where $\langle \psi^2\rangle$ is defined by
$\langle\vec{\pi}^2\rangle=(N-1)\langle \psi^2\rangle$.

The Hartree approximation we adopt here is equivalent to the
Hartree-Fock approximation in the general CJT (Cornwall, Jackiw
and Tomboulis) formalism~\cite{baa}. By the requirement of Hartree
approximation, the one-loop corrections of the quantum field
two-point Green's functions are cancelled by the introduced ''mass
counter term''. For the case of finite $N$, however, the $1/N$
corrections in the Hartree factorization do not include the
contributions from the collision effects of the same order of
$1/N$~\cite{ra}.
 Nevertheless, the factorization method provides a reliable
resummation scheme that allows us to treat the growth of quantum
fluctuations driven by spinodal instabilities self-consistently.

The equation of motion of the background field $\phi$ can be directly
obtained from the term $\chi V'(t)$ in the Lagrangian [Eq. (\ref{hartreeH})].
By means of the tadpole condition $\langle \chi ({\bf x},t) \rangle =0$ and
Eq. (\ref{potentialchi}), we have
%eq10
\begin{equation}
\ddot{\phi} (t)+3H\dot{\phi} (t)+\left[M_{\chi}^2(t)-\lambda \phi^2 (t)
\right]\phi (t)=0 \, . \label{classfieldeq}
\end{equation}
To see how the quantum effects influence the dynamics of the background 
field, it is critical to solve the background field $\phi (t)$ 
self-consistently by including the fluctuations
$\langle\chi^2\rangle (t) $ and $\langle\psi^2\rangle (t)$ in the
mass-squared term $M_{\chi}^2(t)$. 

To find the dynamics of the fluctuation fields, we decompose
$\chi(t)$ and $\vec{\pi}(t)$ in the Fourier basis. In the Heisenberg
picture, one has
%eq11
\begin{eqnarray}
\chi({\bf x}, t)& = & \int\frac{d^3k}{8\pi^3} \, \chi({\bf k}, t)=
\int\frac{d^3k}{8\pi^3} \, \left[ b_{\bf k}f_{\chi, \bf k} (t)  +
b^\dagger_{\bf -k}f^{\star}_{\chi, \bf -k}(t) \right ] e^{i {\bf
k}\cdot {\bf x}},  \nonumber \\
 \pi_{i}({\bf x}, t)& = & \int\frac{d^3k}{8\pi^3} \, \pi_{i}({\bf k},t)
 = \int\frac{d^3k}{8\pi^3} \,  \left[ a_{i \bf k}f_{\pi, \bf k} (t) +
a^\dagger_{i \bf -k}f^{\star}_{\pi, \bf -k}(t)  \right ] e^{i {\bf
k}\cdot {\bf x}}, \label{chipi}
\end{eqnarray}
where $a_{i \bf k}$, $b_{ \bf k}$ and $a^\dagger_{i \bf k}$,
$b^\dagger_{ \bf k}$ are the creation and annihilation operators
which obey the commutation relations $[a_{i{\bf k}},
a^{\dagger}_{j{\bf k}'}]= \delta_{i,j}\delta_{{\bf k, k}'}$ and
$[b_{{\bf k}}, b^{\dagger}_{{\bf k}'}]=
  \delta_{{\bf k, k}'}$.
The equations of the mode functions $f_{\chi, \bf k} (t)$ and
$f_{\pi, \bf k} (t)$ can be found from the Heisenberg field
equations given by
%eq12
\begin{eqnarray}
&& \left[\frac{d^2}{dt^2}+3H\frac{d}{dt}+
\frac{k^2}{a^2}+M_{\chi}^2 (t) \right ]f_{\chi, \bf k}(t)=0, \nonumber \\
&& \left[\frac{d^2}{dt^2}+3H\frac{d}{dt}+
\frac{k^2}{a^2}+M_{\pi}^2 (t) \right ]f_{\pi, \bf k}(t)=0.
\label{modeeq}
\end{eqnarray}

Finally, to close these equations self-consistently, the terms
$\langle \chi^2\rangle (t)$ and $\langle \psi^2\rangle (t)$ in the
mass-squared [Eq. (\ref {selfconmass})] can be determined by
the mode functions as
%eq13
\begin{eqnarray}
\langle \chi^2\rangle (t) &=& \int\frac{d^3k}{8\pi^3} |f_{\chi,
{\bf
k}}(t)|^2 \, , \nonumber \\
\langle \psi^2\rangle (t) &=& \int\frac{d^3k}{8\pi^3} |f_{\pi,
{\bf k}}(t)|^2 \, . \label{selfconeq}
\end{eqnarray}
Obviously, fluctuations $\langle \chi^2\rangle (t)$ and $\langle
\psi^2\rangle (t)$ are created from the self-consistent background
state $\phi (t)$ during the inflation. The nonlinearity, or the
self-interaction, is in fact encoded in the self-consistent solutions
to Eqs. (\ref{classfieldeq})-(\ref{selfconeq}).

\subsection{Renormalization and non-equilibrium equations of motion }

Before proceeding further, one needs to understand the issue of
renormalization regarding the divergences associated with the loop
integrals. The divergences can be determined from the large loop
momentum behavior of the mode functions, which can be found from
the WKB-type solutions to Eq. (\ref{modeeq})~\cite{boyn2}. It turns
out that the self-consistent loop integrals in
Eq.~(\ref{selfconeq}) contain both quadratic and logarithmic
divergences. To get the self-consistent renormalized equations,
one has to subtract the divergences from both the bare mass and
the coupling constant. However, for a weak coupling $\lambda <
10^{-14}$ in a typical inflation model, the logarithmic
subtractions can be neglected. Thus, the mass
renormalization can be simplified as
%eq14
\begin{equation}
m_R^2=m^2 +\frac{\lambda}{2}\left(1+\frac{2}{N}\right) \left
[-\frac{1}{8\pi^2} \frac{\Lambda^2}{a^2} \right ],
\end{equation}
with a negligible renormalization for the coupling constant. As a
result, it leads to the renormalized masses $M_{\chi, R}$ and
$M_{\pi, R}$ which are given by
%eq15
\begin{eqnarray}
M^2_{\chi, R}(t) &=& -m_R^2+\frac{3\lambda}{2}\phi^2 (t)+
  \frac{3\lambda}{2N}\langle\chi^2\rangle_R (t) +
\frac{\lambda}{2}\left(1-\frac{1}{N}\right)\langle \psi^2\rangle_R
(t) \,
,\nonumber \\
M^2_{\pi, R}(t) &=& -m_R^2+\frac{\lambda}{2}\phi^2 (t)+
\frac{\lambda}{2N}\langle\chi^2\rangle_R (t) +
\frac{\lambda}{2}\left(1+\frac{1}{N}\right)\langle \psi^2\rangle_R
(t) \, , \label{massrenorm}
\end{eqnarray}
where
%eq16
\begin{eqnarray}
\langle \chi^2\rangle_R (t) &=& \int^{\Lambda} \frac{d^3k}{8\pi^3}
|f_{\chi, {\bf k}}(t)|^2 -
 \frac{1}{8\pi^2} \frac{\Lambda^2}{a^2} \, , \nonumber \\
\langle \psi^2\rangle_R (t) &=& \int^{\Lambda} \frac{d^3k}{8\pi^3}
|f_{\pi, {\bf k}}(t)|^2 -
 \frac{1}{8\pi^2} \frac{\Lambda^2}{a^2} \, .
  \label{selfconeqrenorm}
\end{eqnarray}
Hereafter, we will drop the subscript $R$, as the integrals of
mode functions are always in terms of the renormalized quantities.

\section{Classical fluctuations of the inflaton}

\subsection{Quantum decoherence}

The quantum-to-classical transition of the quantum fluctuations
during the inflation can be investigated from the Schr\"odinger wave
function approach~\cite{sta,lee}. The Hamiltonian describing
the evolution of quantum wave functions of the fluctuations can be obtained
from Eq.(\ref{hartreeH}). In general, the developments of different
fluctuation modes are separable due to the quadratic feature of the
Hartree factorized Hamiltonian. Let us first consider a mode $ {\bf k}$
of the $\chi$ field. In the Schr\"odinger picture, the initial
(time $t_{0}$) vacuum state is specified by
%eq17
\begin{equation}
b_{ {\bf k}}|t_{0} \rangle_S=0 \, ,
\end{equation}
for all ${\bf k}$. At time $t$, the evolved Schr\"odinger state
$|t \rangle_S$ is given by
%eq18
\begin{equation}
 U( t , t_{0}) \, b_{{\bf k}} \, U^{-1} (t , t_{0})
  |t \rangle_S=0 \, , \label{wavefuneq}
\end{equation}
where $U(t, t_0) =\exp{-(i/\hbar)\int_{t_0}^{t}{\mathcal H}(t)d
t}$. In the coordinate representation, the conjugate momentum can
be written as $ \Pi ({\bf k})= -i\hbar
\partial /
\partial \chi( {\bf -k}) \, , $
and thus, Eq.(\ref{wavefuneq}) reduces to
%eq19
\begin{equation}
\left [\chi ({\bf k})
   -\hbar\gamma_k^{-1}(t)\frac{\partial}{\partial \chi ({\bf -k})}
 \right] \Psi[\chi({\bf k}), t]=0 \, ,
\end{equation}
where
%eq20
\begin{equation}
\gamma_k(t)=\frac{1}{2|f_{\chi, k}(t)|^2}
\left[ 1 - 2iF_{\chi,\bf k}(t) \right]\, ,
\end{equation}
%and
%eq21
\begin{equation}
F_{\chi,\bf k} (t)=\frac{a^3 (t) }{2}\frac{d}{d t}|f_{\chi,\bf k}(t)|^2
\, ,
\end{equation}
and $\Psi[\chi ({\bf k}),t]= \langle \chi ({\bf k}) |t \rangle_S$
is the wave function for the field $\chi (\bf {k}) $. The solution
to the equation can be obtained straightforwardly as
%eq22
\begin{eqnarray}
\Psi[\chi (\bf{k}), t] & = &
  {\cal N}^{1/2}_{\bf k}(t)
 \exp \left ( -\frac{1}{\hbar}\gamma_k (t) |\chi (\bf {k})|^2 \right )
    \\ \nonumber
& = & {\cal N}^{1/2}_{\bf k}(t) \exp \left (
-\frac{1}{\hbar} \frac{|\chi(\bf {k})|^2}{2|f_{\chi,\bf k}(t)|^2}
\left[ 1-i2 F_{\chi,\bf k} (t) \right] \right ) \, ,
\end{eqnarray}
where ${\cal N}^{1/2}_{\bf k}(t)$ is the normalization
coefficient. The wave function of the quantum fluctuations is then
the direct product of the wave functions for all ${\bf k}$ modes
of the $\chi$ field. Notice that for an initial vacuum state which
is a pure state, it will remain the pure state under the unitary
evolution. Therefore, the density matrix in the coordinate
representation becomes the wave function times its complex
conjugate given by
%eq23
\begin{eqnarray}
\prod_{ {\bf k}} \rho(\chi({\bf k}), \bar{\chi } ({\bf k}), t) &=&
\prod_{ {\bf k}} \Psi[\chi ({\bf k}), t] \,
\Psi^{\star}[\bar{\chi} ({\bf k}), t]
\,   \nonumber \\
&=& \prod_{ {\bf k}} {\cal N}_{\bf k}(t) \exp \left\{- \frac{ \chi
(\bf {k}) \chi ({\bf -k}) + \bar{\chi} ({\bf k}) \bar{\chi}
(\bf{-k}) }{2|f_{\chi, \bf k} (t)|^2 } \left[1-i 2F_{\chi,\bf k}
(t)\right] \right \} \, .
\nonumber \\
\end{eqnarray}
The density matrix for each mode $\bf k$ can be decomposed
into the diagonal and off-diagonal elements as follows:
%eq24
\begin{widetext}
\begin{eqnarray}
& \, &  \rho[ \chi ({\bf k}) + \delta ({\bf k}), \bar{\chi} ({\bf
k}) - \delta({\bf k}), t]      \\ \nonumber
   & =&  {\cal N}_{\bf k}(t)
\exp \left\{- \frac{1}{|f_{\chi, \bf k}(t)|^2 } \left[ |\bar{\chi}
({\bf k})|^2 -i 2 F_{\chi,\bf k}(t) \left[\bar{\chi}({\bf
k})\delta ({\bf -k})+\bar{\chi }({\bf -k})\delta ({\bf k})\right]
+|\delta ({\bf k})|^2 \right] \right \} \, .
\end{eqnarray}
\end{widetext}
One can immediately recognize that $F_{\chi,\bf k} (t )$ is the
phase of the off-diagonal elements. A quantum-to-classical
transition is implied if the following condition is satisfied:
%eq25
\begin{equation}
 F_{\chi,\bf k}( t )  \gg
  1\, . \label{quan-to-class}
\end{equation}
Similarly for mode $\pi$, the condition of the quantum-to-classical
transition is given by
%eq26
\begin{equation}
 F_ {\pi,{\bf k}} (t) = \frac{a^3 (t) }{2}
\frac{d}{dt}|f_{\pi, {\bf k}}(t)|^2 \gg 1 .
\end{equation}
Hence, the quantum-to-classical transition is determined by the
behavior of the time-dependence of the mode functions
$f_{\chi, \bf k}(t)$ and $f_{\pi,{\bf k}}(t)$.

To solve Eq. (\ref{modeeq}), we consider the early stage of the
inflation when fluctuations have not grown significantly yet in
the presence of spinodal instabilities as well as the Hubble
parameter remains constant. In this stage, the masses $M_{\chi}
(t)$ and $M_{\pi }(t)$ can be approximated by $-m$ in Eq.
(\ref{massrenorm}). Thus, both mode equations in Eq.
(\ref{modeeq}) can be written as
%eq27
\begin{equation}
\left[\frac{d^2}{dt^2}+3H\frac{d}{dt}+ \frac{k^2}{a^2}-m^2 \right
]f_{\bf k}(t)=0 \, .
\end{equation}
The general solution can be expressed as a combination of the
Hankel function $H_{\nu} (k/aH)$ and the Neumann function $N_{\nu}
(k/aH)$ with $\nu =\sqrt{(9/4) +(m/H)^2}$. In particular, for
fluctuations in superhorizon regime where $k \ll aH $, the growing
mode solution leads to
%eq28
\begin{equation}
f_{\bf k} \simeq f_0 (k) \, e^{(\nu-3/2)Ht}\, ,
\label{approxsolution}
\end{equation}
where $f_0 (k)$ depends on the initial conditions of the mode
functions as we set $t_0 =0$. Apparently, $F_{\bf k} ( t ) \gg 1$
is valid after several e-foldings of inflation when the modes
become superhorizon-sized.

\subsection{Equations of classical fluctuations}

Using the density matrix, one can find the Wigner function from
%eq29
\begin{eqnarray}
W(\bar{\chi} ({\bf k})\, , \, \bar{\pi}_{\chi} ({\bf k}) \, , t)
&=& \int d\left(\frac{\delta }{2\pi}\right)\, e^{-\frac{i}{\hbar}
\bar{\pi}_{\chi}({\bf k}) \delta({\bf k})} \,
 \rho \left( \bar{\chi}({\bf k})-\frac{\delta({\bf k})}{2},\bar{\chi}
({\bf k})+ \frac{\delta({\bf k})}{2}, t \right )  \nonumber \\
 & = & P [|\bar{\chi} ({\bf k})| ]
   \left( N_{\bf k}
  \exp \left[- \frac{ |f_{\chi, \bf k}(t)|^2}{\hbar}
\left |\bar{\pi}_{\chi} ({\bf k})-
   \frac{F_{\chi, \bf k} (t)}{|f_{\chi, \bf k}(t)|^2} \,
\bar{\chi} ({\bf k})\right |^2
 \right ] \right), \nonumber \\ \label{wigner}
\end{eqnarray}
where
%eq30
\begin{equation}
P[|\bar{\chi} ({\bf k})|]=
\left( \frac{1}{\hbar\pi \, |f_{\chi,\bf k}(t) |^2}
 \right)^{\frac{1}{2}} \,
  \exp\left [-\frac{|\bar{\chi} ({\bf k})|^2}{\hbar \, | f_{\chi, \bf
  k}(t)|^2}\right] \, . \label{pdf}
\end{equation}
Since the Wigner function in Eq.(\ref{wigner}) is definite
positive, it can be interpreted as a distribution function in the
phase space of stochastic field perturbations.

For the perturbation modes just crossing out the horizon, the
dynamics of the classical fluctuations can be
described by the Wigner function in the limit of $\hbar
\rightarrow 0$ while keeping $\hbar |f_{\chi, \bf k}(\tau)|^2$
fixed. In fact, the term $\hbar |f_{\chi,\bf k}(\tau)|^2$
basically measures the variance of the field fluctuations
$\bar{\chi} ({\bf k})$ and will grow nonperturbatively due to the
spinodal instabilities. As a result, the Wigner function becomes
%eq31
\begin{equation}
W(\bar{\chi} ({\bf k})\, , \, \bar{\pi}_{\chi} ({\bf k}) \, , t)
=P [|\bar{\chi} ({\bf k})| ] \, \delta \left(  \left
|\bar{\pi}_{\chi} ({\bf k})-
   \frac{F_{\chi,\bf k} (t)}{|f_{\chi, \bf k}(t)|^2} \,
\bar{\chi} ({\bf k})\right
   |\right) \, . \label{semiclasswig}
\end{equation}
The delta function on the above expression yields the equations
of motion for the classical fluctuations at superhorizon scales,
%eq32
\begin{equation}
\bar{\pi}_{\chi} ({\bf k},t)-
   \frac{F_{\chi, \bf k} (t)}{|f_{\chi, \bf k}(t)|^2} \,  \bar{\chi} ({\bf
   k}, t) =0 \, . \label{semiclasseq}
\end{equation}
In terms of the classical fluctuation $\bar{\chi} ({\bf k}, t)$,
Eq. (\ref{semiclasseq}) can be rewritten as
%eq33
\begin{equation}
\frac{ d \bar{\chi}({\bf k},t)}{dt}-
   \frac{1}{2} \left( \frac{d}{d t} \ln  |f_{\chi, \bf k}(t)|^2 \right)
\,  \bar{\chi} ({\bf
   k}, t) =0 \, . \label{stochasticeq}
\end{equation}
With the mode functions $f_{\chi,\bf k}(t)$ derived self-consistently from
Eqs. (11), (14), and (15), Eq. (\ref{stochasticeq}) describes the dynamical
evolution of the mean value of the stochastic fields $\bar{\chi} ({\bf k},t)$
in the superhorizon regime. Obviously, the stationary solution to the
probability distribution function (PDF) of variables  $\bar{\chi}({\bf k},t)$
can be obtained from Eq. (\ref{pdf}).

In contrast to the usual second order evolution equation for the
field $\chi( {\bf k},t)$, Eq. (\ref{stochasticeq}) is a diffusion
type equation which typically occurs  during the overdamped
process of harmonic oscillators. To accommodate the overdamped
behavior of the mean field $\bar{\chi} ({\bf k},t)$ into the mode
equations, the equations for the mode function Eq. (\ref{modeeq})
at superhorizon scales can be further approximated by
%eq34
\begin{equation}
\left[ 3 H \frac{d}{d t} + \frac{k^2}{a^2 (t)} + M_{\chi}^2 (t)
\right] f_{\chi, {\bf k}} (t)= 0 \, . \label{stochastimodeeq}
\end{equation}
As a consequence, Eq. (\ref{stochasticeq}) reduces to
%eq35
\begin{equation}
3H \, \frac{d \bar{\chi}({\bf k}, t)}{dt}  + \left[ \frac{k^2}{a^2(t)}
  + M_{\chi}^2 (t) \right ] \bar{\chi}({\bf k},t)= 0 \, .
\label{stochasticfluctuationchi}
\end{equation}
As expected, these equations are the classical Euler-Lagrange
field equations for the superhorizon modes of fluctuations
$\chi ({\bf k},t)$ in the overdamped regime.

Following the same procedure, we can obtain a similar equation
for the fluctuating field $\pi( {\bf k}, t)$ as
%eq36
\begin{equation}
3H \, \frac{d \bar{\pi_i }({\bf k}, t)}{dt}  + \left[
\frac{k^2}{a^2(t)}
  + M_{\pi}^2 (t) \right ] \bar{\pi_i}({\bf k},t)= 0 \, .
\label{stochasticfluctuationpi}
\end{equation}

\subsection{The PDF of classical fluctuations}

The statistical properties of the classical fluctuation on mode
${\bf k}$ can be characterized by the PDF given by Eq. (\ref{pdf}),
%eq37
\begin{eqnarray}
P[|\pi_{i} ({\bf k})|]&=& \left( \frac{1}{\pi |f_{\pi, {\bf
k}}(t)|^2}\right)^{1/2} \exp\left [-\frac{\pi^2_{i} ({\bf
k})}{| f_{\pi, {\bf k}}(t)|^2}\right] \, , \nonumber \\
P[|\chi ({\bf k})|] &=& \left( \frac{1}{\pi |f_{\chi, {\bf
k}}(t)|^2}\right)^{1/2} \exp\left [-\frac{\chi^2 ({\bf k})}
  {| f_{\chi, {\bf k}}(t)|^2}\right]. \label{pdf1}
\end{eqnarray}
Apparently, the PDFs of the classical perturbation modes
$\pi_{i} ({\bf k})$ and $\chi({\bf k},t)$ are Gaussian with
variances $| f_{\pi, {\bf k}}(t)|^2$  and $| f_{\chi, {\bf k}}(t)|^2$
respectively under the Hartree approximation. It implies that
%eq38
\begin{eqnarray}
\langle \chi^{2n+1}({\bf k}, t) \rangle = 0 \,  ; & \hspace{5mm} &
\langle \pi^{2n+1}_i({\bf k}, t) \rangle = 0 \, , \\ \nonumber
\langle \chi^{2n}({\bf k},t) \rangle = (2n-1)!!
   \langle \chi^2({\bf k},t) \rangle^n \, ;  & \hspace{5mm} &
\langle \pi^{2n}_i({\bf k},t) \rangle = (2n-1)!!
   \langle \pi^2_i({\bf k},t) \rangle^n \, ,
\end{eqnarray}
where $n=1,2,...$. However, the variances generally are  ${\bf
k}$-dependent. Such ${\bf k}$-dependence of variances eventually
leads to non-trivial correlations between fluctuation modes which
will be scrutinized later.

In summary, after a quantum-to-classical transition, the dynamics
of the classical stochastic fields $\bar{\chi} ( {\bf k},t)$,
$\bar{\pi_i} ( {\bf k},t)$ must be based on
Eq. (\ref{stochasticfluctuationchi}) and
Eq. (\ref{stochasticfluctuationpi}) with the evolution of the
self-consistent mode functions. The corresponding PDF of the
perturbations is governed by Eq. (\ref{pdf1}). In fact, by adding a
noise term to the right hand side of both Eqs. (\ref{stochasticfluctuationchi})
and (\ref{stochasticfluctuationpi}), they become Langevin-like equations
which is capable of describing the stochastic properties of perturbations.
Thus, the noise correlations can be determined by the variances of the PDFs.
For the case of single free inflaton, the dynamics of the superhorizon-sized
fluctuations is governed by a Langevin equation~\cite{lee}, which is obtained
by coarse graining the degrees of freedom of all subhorizon modes~\cite{re}.
According to the results we obtained here, the coarse grained Langevin
equation approach is effective not only for the situation of a free scalar
inflaton field, but also for models with self-interaction such as
the $O(N)$ inflation.

Moreover, it should be pointed out that Eqs. (\ref{stochastimodeeq}) -
(\ref{stochasticfluctuationchi}) and their counterparts for the fluctuating
$\pi( {\bf k}, t)$-field are not closed if considering only the classical
fluctuations or the superhorizon  modes. By virtue of $M_{\chi}^2(t)$ and
$M_{\pi}^2(t)$, the subhorizon modes also contribute towards
these diffusion type equations.
This feature is anticipated because of the inevitable
coupling between all  wavelength modes under a substantial
self-interaction. Thus, the renormalized mass squared terms play critical
roles as phenomenological potentials. Consequently,
Eqs. (\ref{stochastimodeeq}) - (\ref{stochasticfluctuationchi}) and their
counterparts for the $\pi( {\bf k}, t)$-field provide a possible scheme
to study the interactions of inflaton by comparing the phenomenological
potentials given by data fitting with the predicted $M_{\chi}^2(t)$
and $M_{\pi}^2(t)$. This characteristics is a main result of this paper.

We now turn our attention to the cosmological implications from
the off-equilibrium dynamics of the $O(N)$ inflationary model  in
the next section.

\section{The Power Spectrum of Primordial Perturbations}

\subsection{The mass density perturbations}

The power spectrum of primordial perturbations is described by
${\cal P}(k)=\langle |\delta_{\bf k}|^2 \rangle$ where the mass
density perturbations are determined by the gauge invariant
quantity~\cite{ko}
%eq39
\begin{equation}
\delta_{\bf k} = \left .\frac{\delta \rho}{\rho+p} \right |_{k=aH}
\, . \label{masspert}
\end{equation}
The mass density fluctuation $\delta \rho$ originates from the
field fluctuations $\pi_{i}$ and $\chi$  given by
%eq40
\begin{eqnarray}
\frac{\delta \rho}{N} &=& \left ( 1- \frac{1}{N}\right )
\left [\frac {1}{2}\langle \dot{\psi}^2\rangle
   + \frac{1}{2a^2}\langle (\nabla \psi)^2\rangle
   -\frac{1}{2}m^2\langle \psi^2\rangle
   +\frac{\lambda}{4}\phi^2\langle \psi^2\rangle
   +\frac{\lambda}{4N}\langle \chi^2\rangle\langle \psi^2\rangle\right. \\ \nonumber
&+&\left. \frac{\lambda}{8}\left(1+\frac{1}{N}\right) \langle \psi^2\rangle^2 \right]
  + \frac{1}{N}\left [\frac{1}{2}\langle\dot{\chi}^2\rangle
 + \frac{1}{2a^2}\langle (\nabla \chi)^2\rangle
   - \frac{1}{2}m^2\langle \chi^2\rangle        \right. \\ \nonumber
&+&\left. \frac{3\lambda}{4}\phi^2\langle \chi^2\rangle
  +\frac{\lambda}{4}\left(1-\frac{1}{N}\right)\langle\psi^2\rangle\langle\chi^2\rangle
  + \frac{3\lambda}{8N}\langle \chi^2\rangle^2 \right ] \, ,
\label{deltadensity}
\end{eqnarray}
with~\cite{boyn2}
%eq41
\begin{eqnarray}
\langle \dot{\psi}^2 \rangle (t)  &=& \int \frac{d^3 {\bf k}}{
(2\pi)^3} |\dot{f}_{\pi, \bf k}(t)|^2 -\left [ \frac{1}{8\pi^2}
\frac{\Lambda^4}{a^4}+ \frac{1}{ 8\pi^2} \frac{\Lambda^2}{a^2}\left(
M^2_{\chi}(t)-\frac{R}{6} +2 \frac{\dot{a}^2}{a^2}\right)\right ]\, ; \nonumber \\
\langle \dot{\chi}^2 \rangle (t) &=& \int \frac{d^3 {\bf k}}{
(2\pi)^3} |\dot{f}_{\chi, \bf k}(t)|^2 -\left [ \frac{1}{8\pi^2}
\frac{\Lambda^4}{a^4}+ \frac{1}{ 8\pi^2}
\frac{\Lambda^2}{a^2}\left( M^2_{\pi}(t)-\frac{R}{6} +2
\frac{\dot{a}^2}{a^2}\right)\right ]     \, ,
\end{eqnarray}
%eq42
\begin{eqnarray}
\langle (\nabla {\psi})^2 \rangle (t)&=& \int \frac{d^3{\bf k}}{
(2\pi)^3} k^2 |f_{\pi, \bf k}(t)|^2- \left [\frac{1}{8\pi^2}
\frac{\Lambda^4}{a^2} - \frac{\Lambda^2}{8\pi^2}
\left(M^2_{\chi}(t)-\frac{R}{6}\right) \right] \, ; \nonumber \\
 \langle (\nabla {\chi})^2 \rangle (t) &=& \int \frac{d^3 {\bf k}}{(2\pi)^3}
k^2 |f_{\chi, \bf k}(t)|^2- \left [ \frac{1}{8\pi^2}
\frac{\Lambda^4}{a^2} - \frac{1\Lambda^2}{8\pi^2}
\left(M^2_{\pi}(t)-\frac{R}{6}\right)  \right] \, ,
\end{eqnarray}
where the Ricci scalar $R=6(\dot{a}^2/a+\ddot{a}/a)=12 H^2$.
Similar to the renormalization of Eq. (\ref{selfconeqrenorm}), we
have ignored the logarithmic subtractions in the renormalization
of above equations. The term summing up the energy
density and the pressure, $\rho+p$, in Eq. (\ref{masspert}) can
be obtained by
%eq43
\begin{eqnarray}
 \frac{\rho+p}{N}=\dot{\phi}^2+
\left ( 1- \frac{1}{N}\right ) \left[\langle\dot{\psi}^2\rangle +
 \frac{1}{a^2}\langle(\nabla\psi)^2\rangle \right]+
\frac{1}{N}\left[\langle\dot{\chi}^2\rangle +
 \frac{1}{a^2}\langle(\nabla\chi)^2\rangle \right ] \, .\label{pressure}
\end{eqnarray}
Apparently, the first term $\dot{\phi}^2$ in Eq. (\ref{pressure})
comes from the background inflaton field. The other terms,
however, are the contributions from the fluctuations.

\subsection{Numerical examples}

We present numerical examples in this section to demonstrate the
self-interaction effect of the $\phi(t)$ field upon the power spectrum.
To begin with, we need to find the self-consistent solutions to the
background field $\phi$ and the mode functions $f_{\chi, \bf k}(t)$ and
$f_{\pi, \bf k}(t)$ from Eqs. (\ref{classfieldeq}) and (\ref{modeeq}).
We assume $\phi (0)\approx 0$ and $\dot{\phi} (0) =0$ for the
background field at the onset of the inflation. The initial $(t=0)$
conditions of the mode functions can be specified as effective free
massive scalar fields in an expanding universe:
%eq44
\begin{eqnarray}
f_{\chi, \bf k} (0) &=& \frac{1}{\sqrt{ 2 [k^2 + M^{2}_{\chi}(0)]}}
\,\, , \,\,\,\,   \dot{f}_{\chi, \bf k} (0) =- i
 \sqrt{\frac{k^2 + M^{2}_{\chi}(0)}{2} } \, ; \nonumber \\
 f_{\pi, \bf k} (0) &=& \frac{1}{ \sqrt{ 2 [k^2 + M^{2}_{\pi}(0)]}}
 \,\, , \,\,\,\,\,    \dot{f}_{\pi, \bf k} (0) =- i
 \sqrt{\frac{k^2 + M^{2}_{\pi}(0)}{ 2 } } \, , \label{initial}
\end{eqnarray}
where we have set $a(0)=1$, and $M^{2}_{\chi}(0)$, $M^{2}_{\pi}(0)$ are the effective
mass square of the field components $\chi$ and $\pi$,
respectively. The values of $M^{2}_{\chi}(0)$ and $M^{2}_{\pi}(0)$
can be zero or a positive number, which depends on the details of
the onset of the inflation. Although this uncertainty really is not
decisive to the evolution of the mode functions $f_{\chi, \bf k}(t)$
and $f_{\pi, \bf k}(t)$, it turns out to become crucial for
correlating the fluctuations as we will see later.

%fig1
\begin{figure}[t]
\begin{center}
\leavevmode
\epsfxsize=5.0in
\epsffile{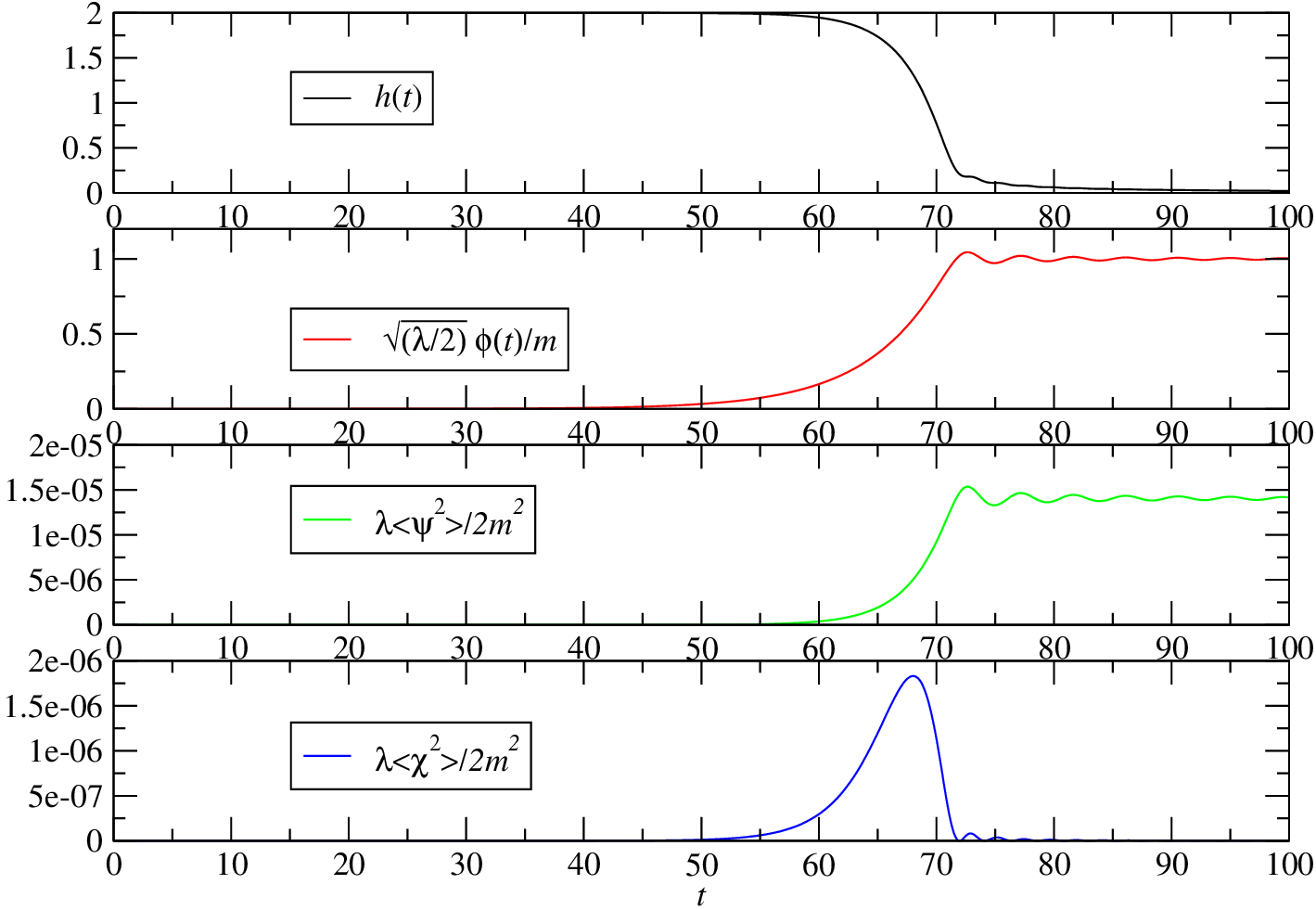}
\end{center}
\caption{The evolution of $h=\frac{H}{m}$,
$\sqrt{\frac{\lambda}{2}} \frac{\phi}{m}$, $\frac{\lambda}{2 m^2}
\langle \psi^2 \rangle$, and $\frac{\lambda}{2 m^2} \langle \chi^2
\rangle$ respectively vs. $t$ (in units of $m^{-1}$) for initial
conditions $ H(0)= 2m $, $\phi(0) \approx 0$, $\dot{\phi}(0)=0$,
$M_{\chi}^2(0)=M_{\pi}^2(0)=m^2$ with $ \lambda =10^{-14}$ in the
case of $N=4$. } \label{fig1}
\end{figure}

Figure 1 plots the expansion rate $h(t)$ of the universe, the 
self-consistent solution to the background field $\phi(t)$, and the 
fluctuations $\langle\psi^2\rangle(t)$, $\langle \chi^2 \rangle (t)$ 
where the parameters are taken to be $\lambda = 10^{-14}$, $N=4$, 
and $H(0)=2 m $, $M^{2}_{\chi}(0)= M^{2}_{\pi}(0)\simeq m^2 > 0$. 
The value for $H(0)$ is determined from the energy density of the 
inflaton field given by Eq. (\ref{hubbleeq}). During the
inflation, the energy density of the inflaton is dominated by that
of the potential energy which leads to $ h^2 (0)= (H/m)^2 \approx N
m^2/M^2_{\rm Pl} \lambda \approx {\cal O} (1)$ for $ N \approx
{\cal O} (1)$, $\lambda \approx 10^{-14}$, and the inflaton mass
$m \approx 10^{12}$ GeV. As a consequence, the Hubble parameter
remains approximately a constant $h\simeq 2$ at $ t\leq 60m^{-1}$,
but drops dramatically after $t\geq 70m^{-1}$. Apparently, $ t\leq
70m^{-1}$ specifies the inflationary epoch.

It has been pointed out earlier that we treat the dynamics of the 
Hubble parameter $H(t)$ classically without any quantum correction
involved in Eq. (\ref{hubbleeq}). It turns out that the background 
field does follow its classical equation of motion as the quantum
corrections in Eq. (\ref{classfieldeq}) give rise to a small
$\delta\rho/\rho \approx 10^{-4}$ in our case. Thus, when the 
semi-classical Einstein equations are employed for a truly 
self-consistent approach, it yields a similar solution of $h$ as 
shown in \cite{boyn2}.

Figure 1 also shows that $\phi(t)\simeq\langle\psi^2\rangle(t)
\simeq\langle\chi^2\rangle(t)\simeq 0$ during the first stage of the 
inflation $t<40m^{-1}$ in the spinodal regime where the initially 
positive $M^{2}_{\chi}$ and $M^{2}_{\pi}$ turn negative quickly. 
The spinodal instabilities become important eventually, and leads 
to the significant growth of both $\langle\psi^2\rangle(t)$ and 
$\langle\chi^2\rangle(t)$ starting at $t= 55m^{-1}$. As expected, 
these properties are not sensitive to the choice of the initial 
effective masses $M^{2}_{\chi}(0)$ and $M^{2}_{\pi}(0)$. It can 
also be seen from the mode function $|f_{\pi,\bf k}(t)|^2$ shown 
in Fig. 2.

Evidently, the mode function undergoes a substantial increase in
the period of $55m^{-1}<t< 70m^{-1}$, but varies much slower after
$t = 70m^{-1}$. This is owing to the predomination of the
criterion $M^2_{\pi} < 0$ in the spinodal regions spawned during
the slow rolling of the inflaton. On the other hand, the spinodal
condition is only weakly satisfied over a span of $70m^{-1}\leq t
\leq 80m^{-1}$ where the background $\phi$ starts the rapid
falling into the valley of the inflaton potential. Therefore, the
process of spinodal instability terminates at time $t_e$ given by
%eq45
\begin{equation}
 M^2_{\pi} (t_e) =
-m^2+\frac{\lambda}{2}\phi^2 (t_e) +
\frac{\lambda}{2N}\langle\chi^2\rangle  (t_e) +
\frac{\lambda}{2}\left (1+\frac{1}{N} \right )\langle \psi^2\rangle
(t_e) \approx 0 \, . \label{endmass}
\end{equation}
It renders $t_e \simeq (70-80) m^{-1}$.

%fig2
\begin{figure}[t]
\begin{center}
\leavevmode
\epsfxsize=5.0in
\epsffile{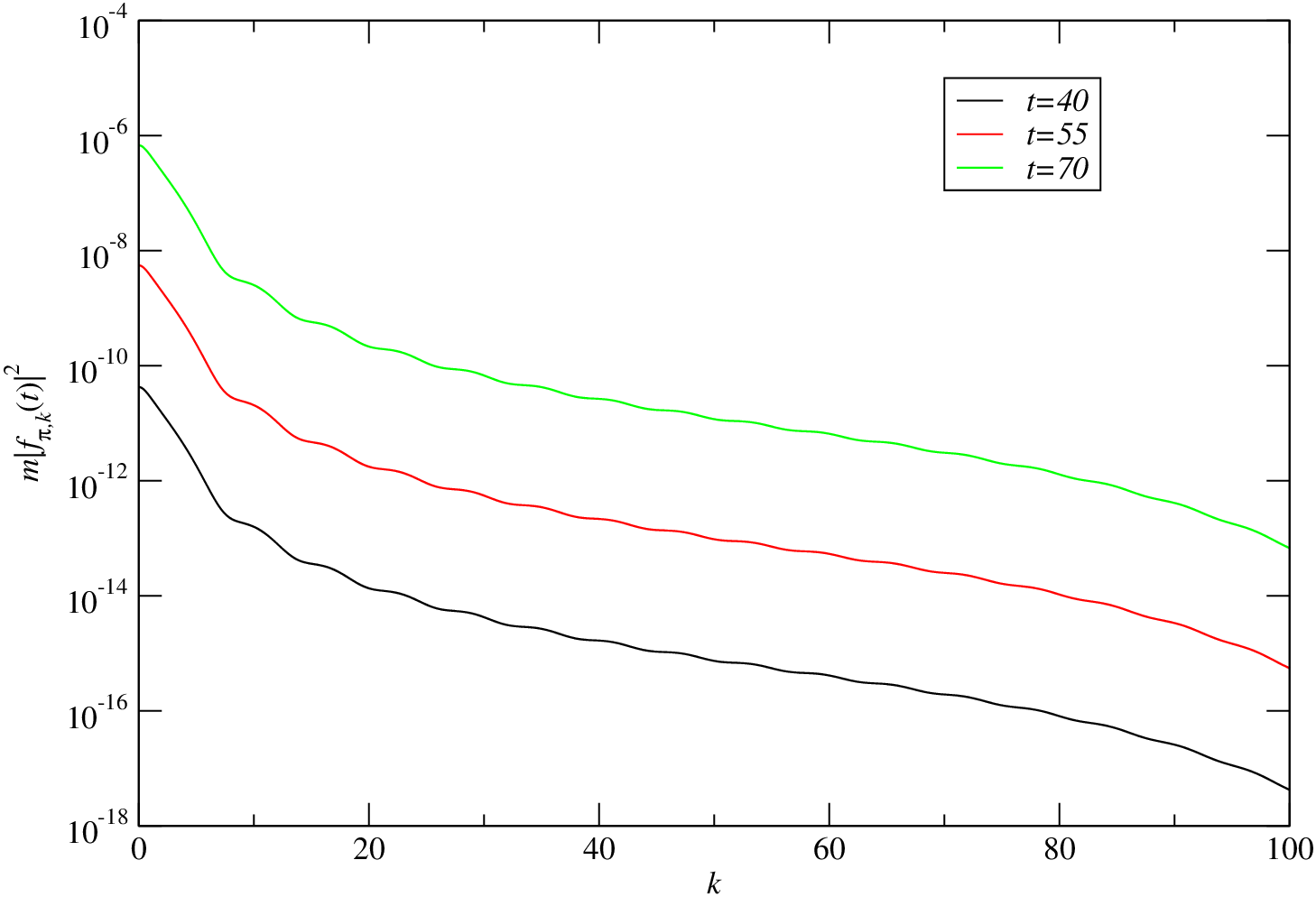}
\end{center}
\caption{Mode function $|f_{\pi,\bf k} (t)|^2$ vs. $k $ (in units of
$m$) at $t=40,~55,~70$ (in units of $m^{-1}$) under the same initial
conditions and parameters as in Fig. 1 for the case of $N=4$. }
\label{fig2}
\end{figure}
The sharp decline of $h$ signifies the end of inflation.
Therefore, the number of the total inflationary e-foldings is
$N_{\rm e} \approx Ht=2m t \approx 140$. The perturbation just
about to enter the present Hubble horizon should be transgressing
the horizon in the inflationary era around $t=t_{60}\simeq
40m^{-1}$, some 60 e-folds before the end of inflation. We find
that  most of energy density of the background $\phi$ state is
transferred to the fluctuations $\langle\psi^2\rangle$ via the
production of $\pi$ modes during the inflationary epoch. Since
$M^2_{\pi} (t_e)\simeq 0$, these $\pi$ modes are in fact the
massless Goldstone modes of the broken symmetry phase. The
Goldstone theorem is fulfilled dynamically.

%fig3
\begin{figure}[t]
\begin{center}
\leavevmode
\epsfxsize=5.0in
\epsffile{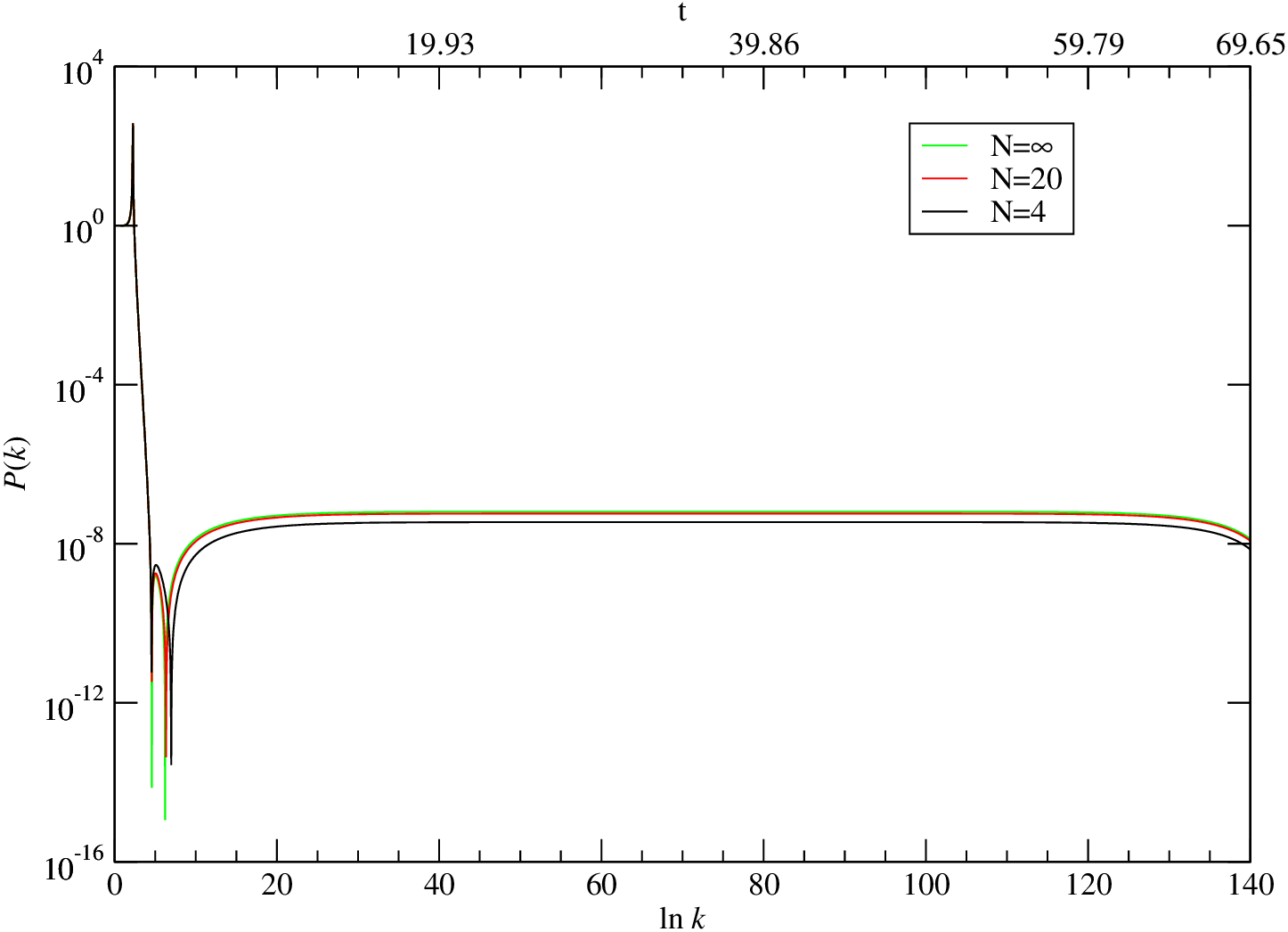}
\end{center}
\caption{The power spectrum of primordial density perturbations
vs. $\ln k$ ($k$ in units of $m$ ) with the same initial conditions
and parameters as in Fig. 1 for the case of $N=4, 20$, and $ \infty$
respectively.} \label{fig3}
\end{figure}

%fig4
\begin{figure}[ht]
\begin{center}
\leavevmode
\epsfxsize=5.0in
\epsffile{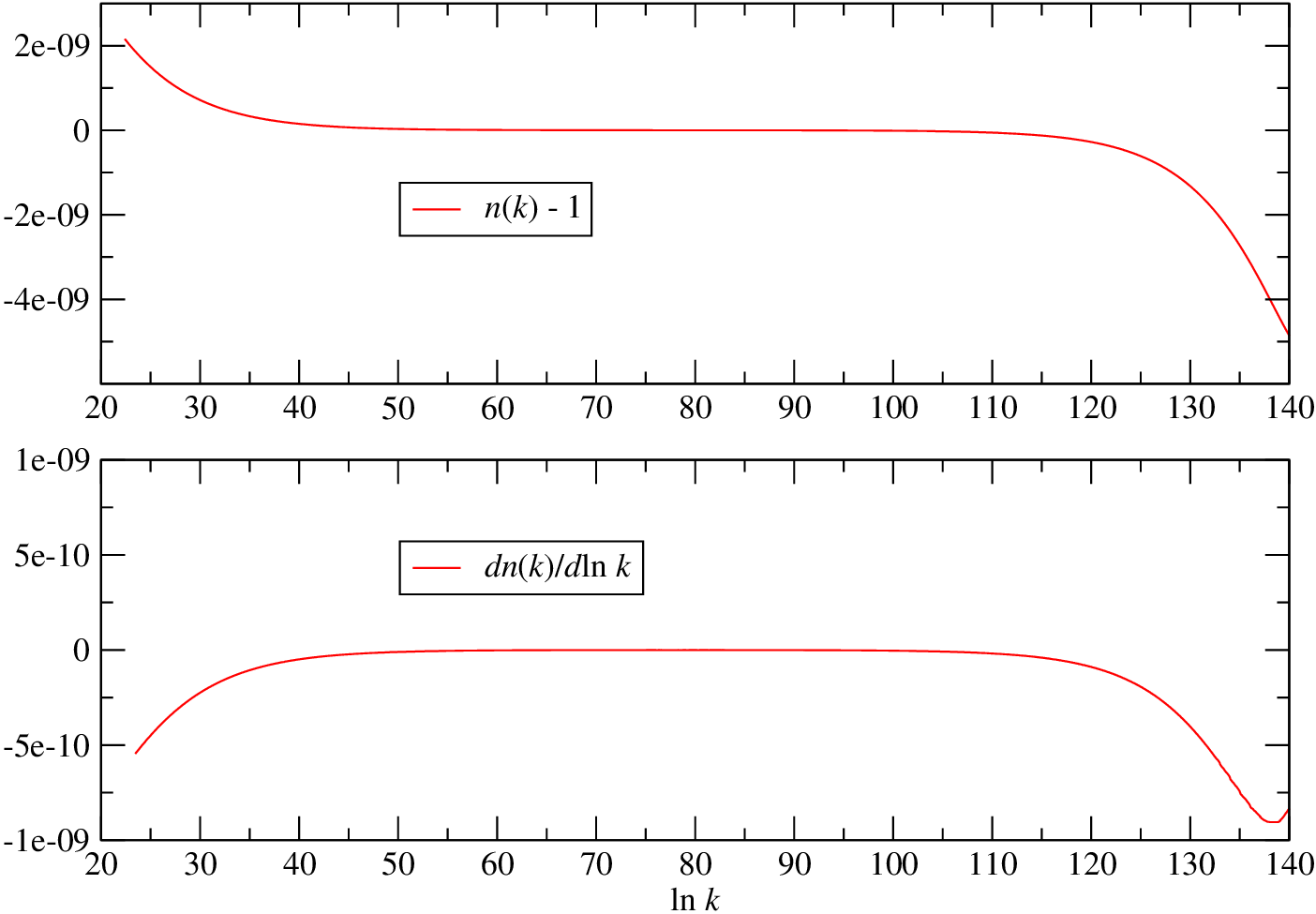}
\end{center}
\caption{Spectral index $n(k)$ and its running $dn(k)/d\ln k$ vs.
$\ln k$ ($k$ in units of $m$) with the same initial conditions
and parameters as in Fig. 1 for the case of $N=4$.}
\label{fig4}
\end{figure}

Using the numerical solutions, we calculate the power spectrum of
primordial perturbations. The result is shown in Fig. 3, in which
$N$ is taken to be 4, 20, and $\infty$. The $N$ dependence of the
power spectrum is trivial. We find that the power is generally
$k$-independent for modes $k$ which cross the horizon when
$t<55m^{-1}$. While the  energy transfer from the background
field to the fluctuations driven by the spinodal instability leads
to a swift increase in $\left\langle\psi^2\right\rangle$, the
inflaton field at later times  rolls down the potential hill which
also results in an increase of $\dot\phi$. This in turn renders the
gradual decrease in power for shorter wavelength modes crossing
out of the Hubble horizon at $t>55m^{-1}$ as the increase of
$\dot\phi$ dominates.

The spectral index $n(k)-1 = dP(k)/d\ln k$ and its $k$-dependence
$dn(k)/d\ln k$ are given in Fig. 4. The value of $n(k)$ varies
from unity at the larger scales as $\ln k < 110$ to about $n< 1$
at the smaller scale as $\ln k > 110$. This leads to a index-running
$dn(k)/d\ln k< 0$ in the wavelength range corresponding to the
horizon-crossing times during $55< mt <70$. Once again, the
physical reason for the running of the spectral index is  due to
the energy transfer from the inflationary background field to the
fluctuations as well as the evolution of the inflaton field.
Although one may refer this energy transfer to the third or even
higher order derivative of the effective potential $V''' (\phi)$
of the inflaton, the index-running shown in Fig.4 can only be
obtained through a proper off-equilibrium dynamics, but not by the
classical effective potential approach.

\section{The correlation of primordial perturbations}

\subsection{Correlation function and initial condition of classical field
fluctuations}

The equal-time two-point correlation functions of fluctuations
have been extensively used to study the  applications of the
$O(N)$ model to the off-equilibrium chiral phase transition in
relativistic heavy ion collisions where the coupling constant is of order
one~\cite{coo}. The correlation caused by the off-equilibrium process
is generic, because the evolution of fluctuations is typically with
respect to a time-dependent background. In our case, the mode
functions of fluctuations $f_{\chi,\bf k}(t)$ and $f_{\pi,\bf k}(t)$ are
not trivial plane waves, they depend rather on the dynamics of
interaction or of self-interaction. Hence, it would be interesting to
search for the correlations of the primordial perturbations.

By means of Eq. (\ref{chipi}), the two-point correlation functions
of the fluctuations are given by
%eq46
\begin{eqnarray}
\langle \chi({\bf x},t)\chi({\bf x'},t) \rangle & = &
  \int \frac{d^3k}{(2\pi)^3}e^{i{\bf k}\cdot ({\bf x-x'})}
   |f_{\chi,\bf k}(t)|^2   \nonumber \\
  & = & \frac{1}{2\pi^2|{\bf x-x}'|}
    \int^{\Lambda} \, k \, dk \, \sin k|{\bf x-x}'||f_{\chi \bf k}(t)|^2
    \, ,
     \nonumber \\
 \langle \pi_i({\bf x},t)\pi_i({\bf x'},t) \rangle & = &
  \int \frac{d^3k}{(2\pi)^3}e^{i{\bf k}\cdot ({\bf x-x'})}
   |f_{\pi,\bf k}(t)|^2  \nonumber \\
  &  = & \frac{1}{2\pi^2|{\bf x-x}'|}
    \int^{\Lambda}\, k \, dk \, \sin k|{\bf x-x}'||f_{\pi \bf k}(t)|^2,
       \label{correlation}
\end{eqnarray}
where we have considered that the dynamics of $|f_{\pi,\bf k}(t)|^2$ and
$|f_{\chi,\bf k}(t)|^2$ is isotropic in ${\bf k}$-space.

It can be seen from Eq. (\ref{correlation}) that both correlation
functions $\langle \chi({\bf x},t)\chi({\bf x'},t) \rangle$ and
$\langle \pi_i({\bf x},t)\pi_i({\bf x'},t) \rangle$ over large spatial
distances $|{\bf x-x}'|$ are sensitive to the long wavelength behavior
of $|f_{\chi,\bf k}(t)|^{2}$ and $|f_{\pi,\bf k}(t)|^2$.
%fig5
\begin{figure}[t]
\begin{center}
\leavevmode
\epsfxsize=5.0in
\epsffile{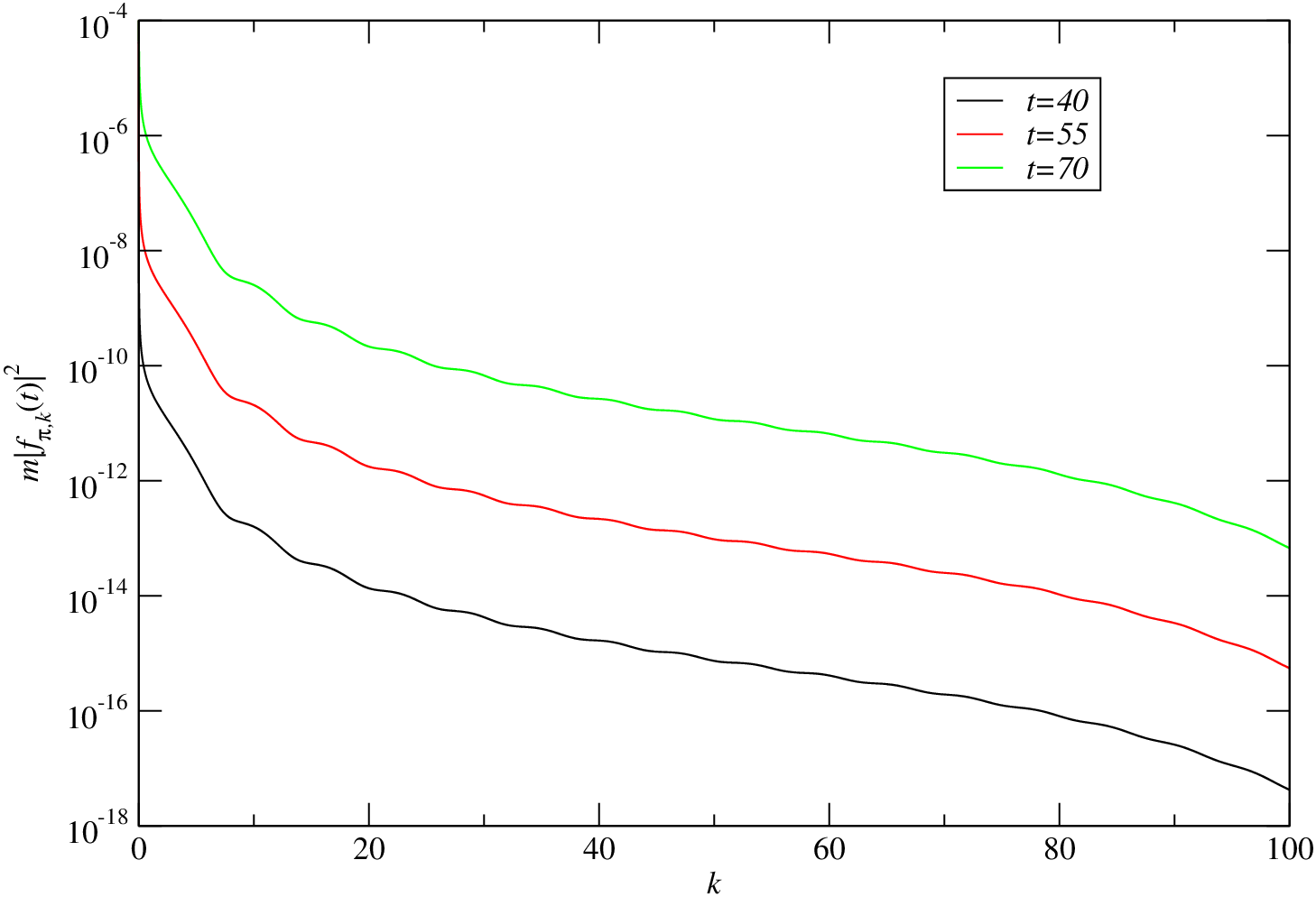}
\caption{Mode function $|f_{\pi, \bf k}(t)|^2$ vs. $k (m)$ at
$t=40,\ 55,\ 70 $ in units of $m^{-1}$. All parameters are specified
the same as in Fig. 2 except the initial conditions Eq. (44) where
$M^2_{\chi}(0)=M^2_{\pi}(0)=0$ are assumed.
}
\label{fig5}
\end{center}
\end{figure}
However, the infrared behavior of the mode functions during the
early stage of the inflation are actually governed by the initial
conditions of the fluctuations as shown in Eq. (\ref{approxsolution}).
For instance, if we take the initial values
$M^2_{\chi}(0)= M^2_{\pi}(0)\simeq m^2 $ in Eq. (\ref{initial}),
then one can show that with Eq. (\ref{approxsolution}) the
correlations decay exponentially as $\exp\{-(m|{\bf x-x'}|)\}=
\exp\{-(H|{\bf x-x'}|/2)\}$ when we set $ H=2m$ to mimic the
constant expansion rate during the inflation. Hence,
the correlation between the inflationary perturbation modes is negligible if the spatial distance
between the modes is larger than a horizon size $a|{\bf x-x'}| >
H^{-1}= 0.5m^{-1}$. On the other hand, if the initial conditions
for the mode functions are chosen as $M^2_{\chi}(0)=M^2_{\pi}(0)=0 $,
then the solution to the mode function can be approximated by
%eq47
\begin{equation}
|f_{\bf k}(t)|^2 \simeq  \frac{(\nu+\frac{3}{2})^2}{k} \,
e^{2[\nu-(3/2)]Ht}.
\end{equation}
This function is singular as $1/k$ when $k \rightarrow 0$. For such
infrared behavior of the mode function, we have
%eq48
\begin{equation}
\frac{\langle \chi({\bf x},t)\chi({\bf x'},t) \rangle} {\langle
\chi^2 \rangle}  \simeq \frac{\langle \pi({\bf x},t)\pi({\bf
x'},t) \rangle} {\langle \pi^2 \rangle}  \propto  \frac{1}{(Ha|{\bf
x-x}'|)^2}. \label{powerlaw}
\end{equation}
This correlation clearly covers spatial range much larger than
the horizon size $H^{-1}$.

\begin{figure}[t]
\begin{center}
\leavevmode
\epsfxsize=5.0in
\epsffile{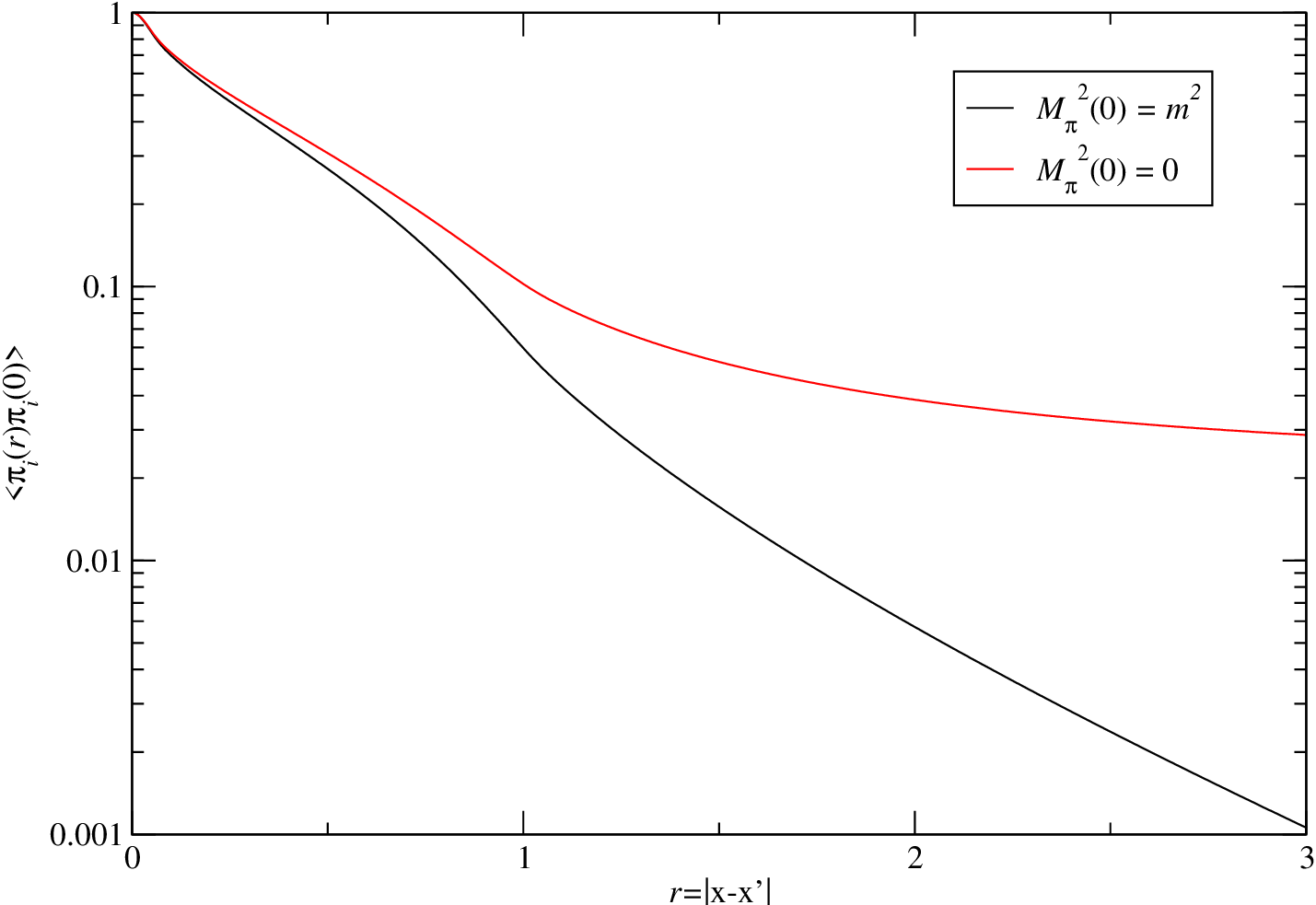}
\caption{$\langle\pi({\bf x},t)\pi({\bf x'},t)\rangle$ vs.
$r=|{\bf x- x'}| $ with initial conditions $M^2_{\pi}(0)=m^2$ (black)
and zero (red) for modes crossing the horizon at $t=40$ $m^{-1}$.
The separation $r$ is in unit of $m^{-1}$.}
\label{fig6}
\end{center}
\end{figure}

The numerical solutions of $|f_{\pi,\bf k}(t)|^2$ with the vanishing
initial mass terms are plotted in Fig. \ref{fig5}. Obviously the
mode functions given in Fig. \ref{fig5} have shown very
different infrared behavior from those in Fig. \ref{fig2}, where
we have taken $M^2_{\chi}(0)=M^2_{\pi}(0)\simeq m^2$. However,
we expect that the tail of the correlation function depends sensitively
on the initial conditions of the inflaton fluctuations. To illustrate
this feature, we plot in Fig. \ref{fig6} the correlation functions
$\langle\pi({\bf x},t)\pi({\bf x'},t)\rangle$ obtained by two different
initial conditions: $M^2_{\pi}(0) > 0$ and $M^2_{\pi}(0)=0$. We see that
the respective behaviors of $\langle\pi({\bf x},t)\pi({\bf x'},t)\rangle$
at scales $r=|{\bf x-x'}|$ larger than 0.5 $m^{-1}$ (which equals the
horizon distance $1/H$) are quite distinguishable. With $M^2_{\pi}=0$
initially, the correlation function scales as a power law:
%eq49
\begin{equation}
\langle \pi_i({\bf x},t)\pi_i({\bf x'},t) \rangle \propto
|{\bf x-x}'|^{-1.8}. \label{index}
\end{equation}
On the other hand, the correlation function clearly possesses an
exponential fall-off with distance when the initial condition is
taken to be $ M^2_{\pi}(0)>0$. This example shows that the initial
conditions can be recalled from the tail of the correlation
function of classical perturbations despite that the original
quantum fluctuations for those superhorizon modes have went
through a quantum-to-classical transition during the
off-equilibrium evolution.

%fig6

\subsection{Detection of the two-point space-scale correlation}

The primordial density perturbations is governed
by the fluctuations of the fields $\chi$ and $\pi_{i}$. Thus,
the field correlation functions $\langle \chi({\bf x},t)\chi({\bf
x'},t) \rangle$ and $\langle \pi_i({\bf x},t)\pi_i({\bf x'},t)
\rangle$ will inevitably lead to the correlation of the
primordial density perturbation $\langle \delta({\bf x},
t)\delta({\bf x'},t)\rangle $.
When perturbations of scale $k$ cross the horizon
at a time $t$, the equal-time two point correlation function
between $({\bf x}, t)$-$({\bf x'},t)$ will yield the correlation
between two space-scale modes $({\bf x}, k)$-$({\bf x'}, k)$
via the mapping formula $k=a(t)H$ [Eq. (\ref{masspert})]. The
density contrast $\delta_{\bf k}({\bf x})$ represents the fluctuations
in total energy density at the spatial point ${\bf x}$ with a scale ${\bf k}$.

Obviously, one cannot measure the density (energy) perturbations
precisely on a scale ${\bf k}$ and at a spatial point ${\bf x}$
simultaneously. The essence of $\delta_{\bf k}({\bf x})$ is as follows.
When a perturbation crosses the horizon at the scale $k=aH$, the
position of the perturbation has an uncertainty typically given by
the size of the horizon, i.e. $\Delta x = (aH)^{-1}$. By virtue of
the uncertainty relation $\Delta x \Delta k \simeq 2\pi$, the scale of
the perturbation thus lies within the band from $k -(1/2)\Delta k$
to $k + (1/2)\Delta k$ where $\Delta k \simeq aH$. Therefore,
$\delta_{\bf k}({\bf x})$ describes the perturbation in a
phase-space cell $({\bf x,k})$ with a size confined by the spatial
range from ${\bf x}$ to ${\bf x}+ \Delta {\bf x}$, and by the scale
range from ${\bf k}$ to ${\bf k} + \Delta {\bf k}$. The
volume of this phase-space cell is characterized by $\Delta {\bf
x}\cdot \Delta {\bf k}\simeq 2\pi$. Consequently,
$\langle\delta_{\bf k}({\bf x})\delta_{\bf k}({\bf x'})\rangle$
governs the correlation between two perturbation modes localized
within the cells centering at $({\bf x,k})$ and $({\bf x',k})$
in the phase ({\bf x-k}) space.  In order to unveil the effects of
$\langle \chi({\bf x},t)\chi({\bf x'},t)\rangle$
and $\langle\pi({\bf x},t)\pi({\bf x'},t)\rangle$, it is
indispensable to decompose the mass density perturbations into
the {\bf x}-{\bf k} modes in the phase space. The discrete wavelet
transform (DWT) is designed to do such space-scale [{\bf x}-{\bf k}]
decomposition~\cite{dau}.

In the formulation of DWT, there are two sets of spatially localized
bases given by the scaling functions $|{\bf j,l}\rangle_s$, and the
wavelet functions $|{\bf j,l}\rangle_{w}$; both are characterized by
the indices ${\bf j}$ and ${\bf l}$. For a 1-D sample with a spatial
size $L$, the index
$j=0, 1, 2,...$ stands for a scale from $k_j$ to $k_j +\Delta k_j
$ in which $k_j =2\pi2^j/L$ and $\Delta k_j =2\pi2^j/L$.
The index $l=0, 1,...,2^j-1$ denotes the location of the spatial
point within $Ll/2^j <x_l < L(l+1)/2^j$. These bases
are complete, and they satisfy the orthogonal relations
${}_s\langle { j, l'}|{ j,l}\rangle_s=\delta_{ l,l'}$, and
${}_w\langle { j',l'}|{ j,l}\rangle_w=\delta_{ j,j'}\delta_{ l,l'}$.
For 3-D samples, the DWT bases are given by the direct product of
the 1-D bases. Thus, a density field $|\delta\rangle$ can be
decomposed into the phase-space modes $({\bf j,l})$ as
%eq50
\begin{eqnarray}
\delta_{\bf j,l}\equiv \ _{s}\langle {\bf j,l}|\delta\rangle
 & = & \int  d^3x \, \delta({\bf x})\phi_{\bf j,l}({\bf x})
  =\int  \frac{d^3k}{(2\pi)^3} \, \hat{\delta}({\bf k})
\hat{\phi}_{\bf j,l}({\bf k})\, ,
  \nonumber \\
\tilde{\delta}_{\bf j,l}\equiv \ _{w}\langle {\bf j,l}|\delta\rangle
 & = & \int d^3 x \, \delta({\bf x})\psi_{\bf j,l}({\bf x})
 =\int  \frac{d^3k}{(2\pi)^3} \, \delta({\bf k})\hat{\psi}_{\bf j,l}({\bf
 k})\, ,
\end{eqnarray}
where $\phi_{\bf j,l}({\bf x})=\langle x|{\bf j,l}\rangle_s$,
$\hat{\phi}_{\bf j,l}({\bf k})=\langle k|{\bf j,l}\rangle_s$,
$\psi_{\bf j,l}({\bf x})=\langle x|{\bf j,l}\rangle_w$
and $\hat{\phi}_{\bf j,l}({\bf k})=\langle k|{\bf j,l}\rangle_w$, i.e.
they are the scaling functions and the wavelet functions in either the
$x$-representation or the $k$-representation, respectively.

Subsequently, the two-point correlation functions in Eq.
(\ref{correlation}) can be rewritten in terms of the DWT bases as
%eq51
\begin{eqnarray}
\langle \tilde {\chi}_{\bf j,l}\tilde{\chi}_{\bf j,l'}\rangle
 &=&  \int \frac{d^3k}{(2\pi)^3} \, |f_{\chi,\bf k}(t)|^2
\hat{\psi}_{\bf j,l}({\bf k})\hat{\psi}^*_{\bf j,l'}({\bf k}) \, ,
\nonumber \\
\langle \chi_{\bf j,l}\chi_{\bf j,l'}\rangle
 &=&  \int \frac{d^3k}{(2\pi)^3} \, |f_{\chi,\bf k}(t)|^2
\hat{\phi}_{\bf j,l}({\bf k})\hat{\phi}^*_{\bf j,l'}({\bf k}) \, ;
\nonumber \\
\langle \tilde{\pi}_{i ; \bf  j,l}\tilde{\pi}_{i; \bf j,l'}\rangle
 &=& \int \frac{d^3k}{(2\pi)^3}\, |f_{\pi,\bf k}(t)|^2
\hat{\psi}_{\bf j,l}({\bf k})\hat{\psi}^*_{\bf j,l'}({\bf k}) \, ,
\nonumber \\
\langle \pi_{i ; \bf  j,l}\pi_{i; \bf j,l'}\rangle
 &=& \int \frac{d^3k}{(2\pi)^3}\, |f_{\pi,\bf k}(t)|^2
\hat{\phi}_{\bf j,l}({\bf k})\hat{\phi}^*_{\bf j,l'}({\bf k}) \, ,
\label{DWTcorrelation}
\end{eqnarray}
where the time $t$ is taken to be $t_j$ specified by the
relation $2\pi 2^j/L=k=aH$. Since $a =\exp(Ht) $, one has
%eq52
\begin{equation}
t_j=\frac{1}{H}\left [j\ln 2 + \ln \left(\frac{2\pi}{LH}\right
)\right ] \, .
\label{tjrelation}
\end{equation}
Thus, Eq. (\ref{DWTcorrelation}) can be used to determine the
correlations between fluctuations at different spatial points
${\bf l}$ and ${\bf l'}$, both crossing out of the Hubble horizon
at the same time $t_j$ during the inflationary epoch.

For example, in the case of a free scalar field $\chi$ with mass
$m$, the DWT mode-mode correlations can be expressed as
%eq53
\begin{eqnarray}
\langle \tilde{\chi}_{\bf j,l}\tilde{\chi}_{i; \bf j,l'}\rangle
 = \int \frac{d^3k}{(2\pi)^3 \, 2 \, (k^2+m^2)^{1/2} }\,
\hat{\psi}^*_{\bf j,l}({\bf k})\hat{\psi}^*_{\bf j,l'}({\bf k}) \, ,
  \nonumber \\
\langle \chi_{\bf j,l}\chi_{i; \bf j,l'}\rangle
 = \int \frac{d^3k}{(2\pi)^3 \, 2 \, (k^2+m^2)^{1/2} }\,
\hat{\phi}^*_{\bf j,l}({\bf k})\hat{\phi}^*_{\bf j,l'}({\bf k}) \,
.
\end{eqnarray}
Since $\hat{\psi}_{\bf j,l}({\bf k})$ is localized at $k\simeq 2\pi 2^j/L$,
and $\hat{\phi}^*_{\bf j,l}({\bf k})$ is non-zero at $k \leq  2\pi 2^j/L$,
one can approximate $(k^2+m^2)^{-1/2}\approx m^{-1} $ if the ${\bf
j}$-scales are greater than the Compton wavelength of mass $m$. Thus,
%eq54
\begin{eqnarray}
\langle \tilde{\chi}_{\bf j,l}\tilde{\chi}_{\bf j,l'}\rangle
 \simeq  \frac{1}{(2\pi)^3 \,  2m} \int \, d^3k \,
\hat{\psi}^*_{\bf j,l}({\bf k})\hat{\psi}^*_{\bf j,l'}({\bf k})= 0
\hspace{4mm} \ {\rm \ if \ } l\neq l' \, , \nonumber \\
\langle \chi_{\bf j,l}\chi_{\bf j,l'}\rangle
 \simeq  \frac{1}{(2\pi)^3 \,  2m} \int \, d^3k \,
\hat{\phi}^*_{\bf j,l}({\bf k})\hat{\phi}^*_{\bf j,l'}({\bf k})= 0
\hspace{4mm} \ {\rm \ if \ } l\neq l' \, ,
\end{eqnarray}
where the orthogonality of the DWT bases with respect to the
indices ${\bf l, l'}$ has been applied. Similar results hold for
the field $\pi_{i; {\bf j,l}}$. Hence, for the free field case the
DWT modes are not correlated and therefore, the correlation
function  $\langle \delta_{\bf k}({\bf x})\delta_{\bf k}({\bf
x'})\rangle$ is trivial, as long as the scale under consideration
is larger than the Compton wavelength of the scalar field.

%fig7
\begin{figure}[t]
\begin{center}
\leavevmode \epsfxsize=5.0in \epsffile{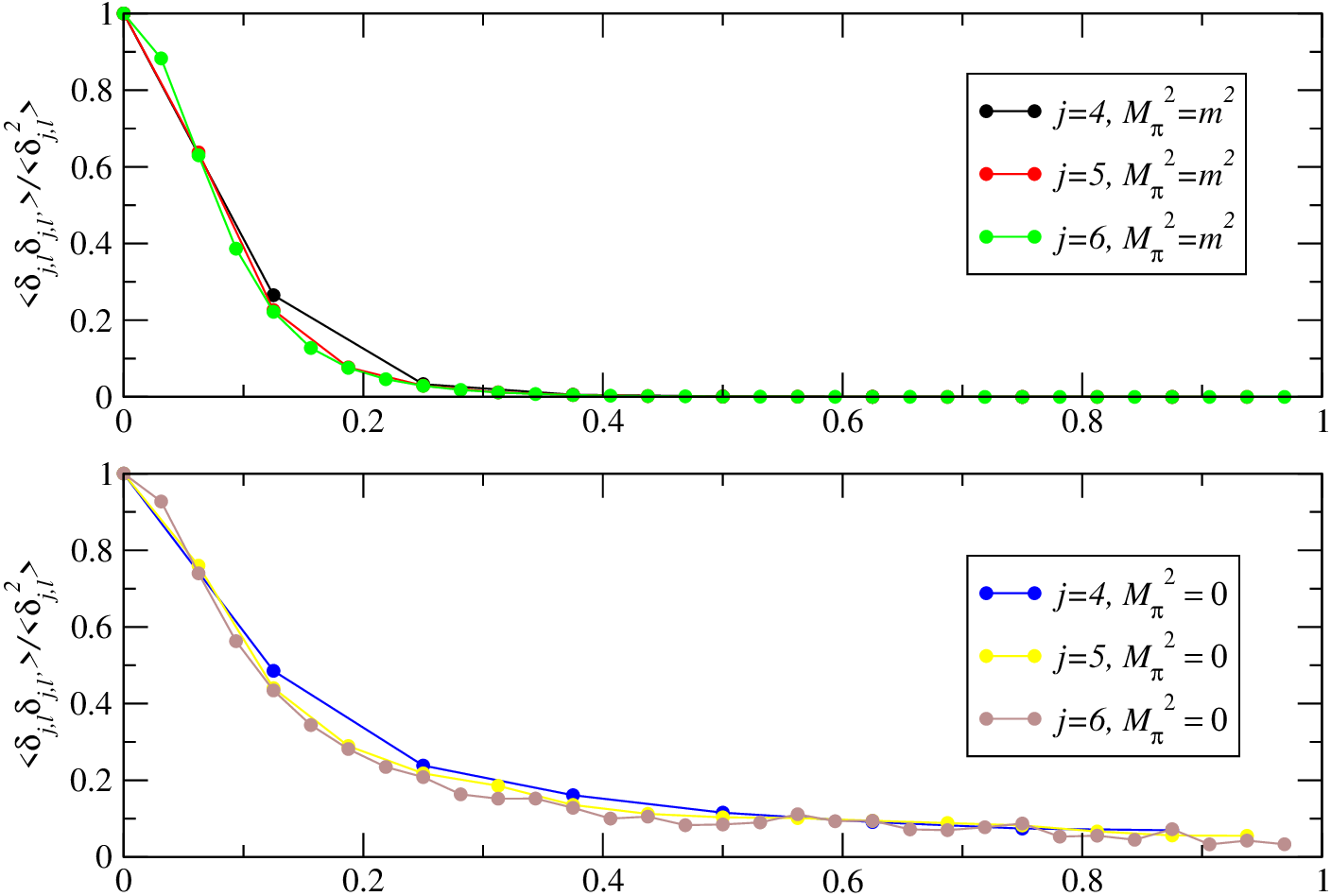} \caption{The
normalized mode-mode correlation functions of density fields in
phase space are plotted against the separation $r$ (in units of
$m^{-1}$) of two perturbation modes with respect to various
scales $j$ under two different sets of initial conditions. The
parameter $L$ is taken to be $2m^{-1}$. } \label{fig7}
\end{center}
\end{figure}

However, for the self-interacting $O(N)$ model the DWT mode-mode
correlations [Eq. (\ref{DWTcorrelation})] are generally non-zero.
Accordingly, the two-point space-scale correlation
$\langle \delta_{\bf k}({\bf x})\delta_{\bf k}({\bf x'})\rangle$
may become non-trivial and initial condition dependent. Using Eqs.
(\ref{masspert})-(\ref{pressure}), it is straightforward to
calculate the normalized two-point space-scale correlation
function defined as
$\langle\delta_{\bf j,l}\delta_{\bf j,l'}\rangle/
\langle\delta^2_{\bf j,l}\rangle$.

As a numerical example, the normalized two-point space-scale
correlations of density perturbations under different initial
conditions are plotted in Fig.~\ref{fig7}, in which the parameters are
taken to be $H=2m$ and $L=2m^{-1}$. From Eq. (\ref{tjrelation}),
one has
%eq55
\begin{equation}
t_j \simeq (0.35 j +0.23) m^{-1}.
\label{tjfig7}
\end{equation}
Thus for $j=4 - 6$ shown in Fig.~\ref{fig7}, we have
$t_j \simeq (1.63 - 2.33)m^{-1}$ which corresponds to the number of
e-foldings $Ht_j\simeq 3.3 - 4.7$. Since the inflation under
consideration lasts from $t=0$ to about $ t= 70 m^{-1}$, the
correlations in Fig.~\ref{fig7} actually probe the inflaton dynamics
at the beginning of the inflation.
With $M^2_{\chi}(0)=M^2_{\pi}(0)\simeq m^2>0$,
the space-scale correlation function approaches zero drastically as
the distance $r$ between the two modes increases; while the correlation
deviates from zero and lasts for a long range for perturbations with
$M^2_{\chi}=M^2_{\pi}=0$ initially.

\begin{figure}[t]
\begin{center}
\leavevmode \epsfxsize=5.0in \epsffile{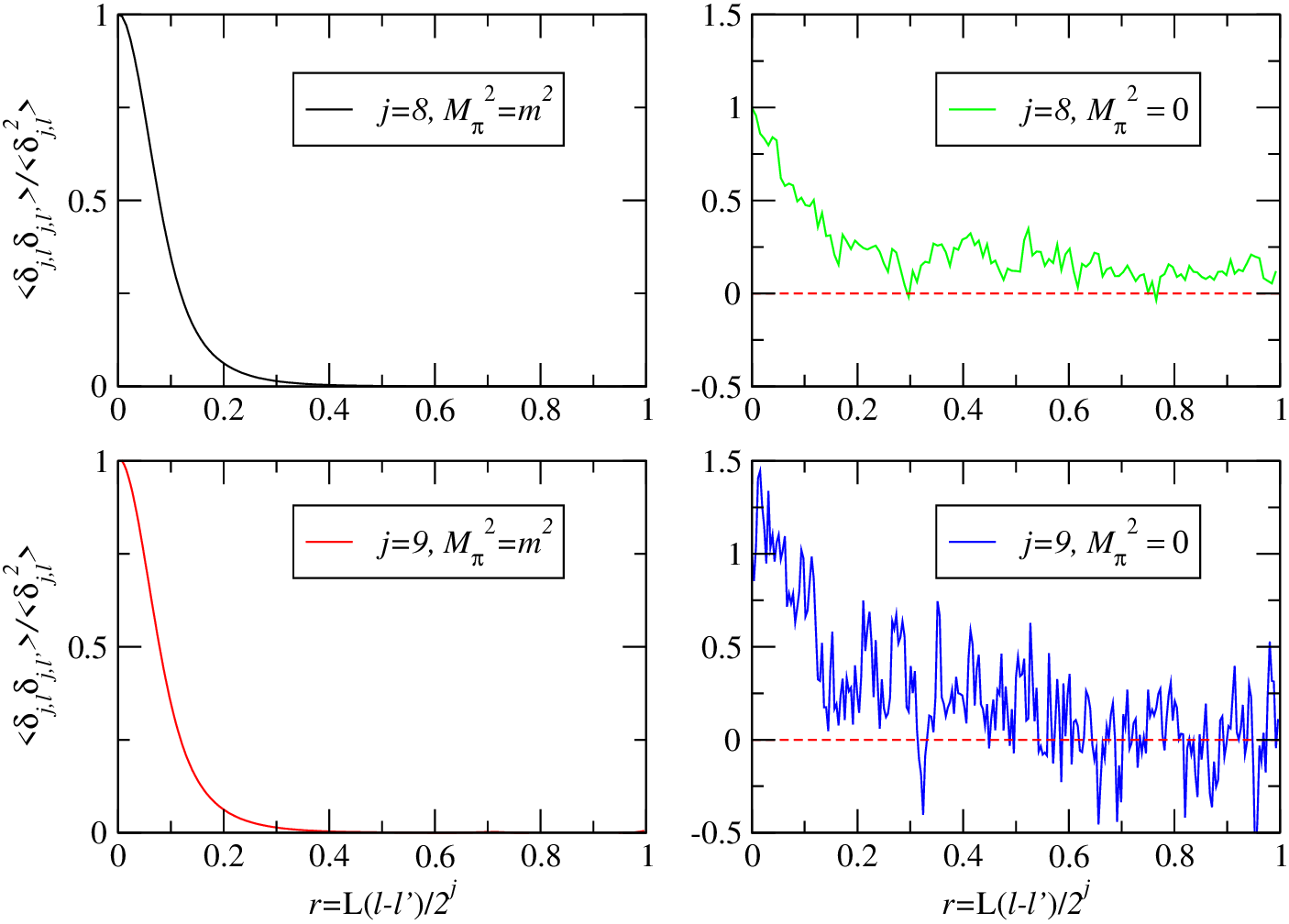} \caption{The
space-scale correlations as a function of $r$ (in units of
$m^{-1}$) under two sets of initial conditions as those in Fig.
7 are plotted with respect to finer scales $j = 8,\ 9$.}
\label{fig8}
\end{center}
\end{figure}

Figure~\ref{fig8} plots the same correlations as in
Fig.~\ref{fig7} but with respect to $j= 8, 9$, or $t_j\simeq (2.75
- 3.38)m^{-1}$, which corresponds to the number of e-foldings
$Ht_j\simeq 5.52 - 6.8$. The behavior of correlations under the
initial condition $M^2_{\chi}(0)=M^2_{\pi}(0)\simeq m^2$ remains
the same. However, the correlations with
$M^2_{\chi}(0)=M^2_{\pi}(0)= 0$ do not coincide with those shown
in Figure~\ref{fig7}. That is, the space-scale correlation of
density perturbations in the model with $M^2_{\pi}>0$ initially
does not experience a significant evolution during the first few
e-foldings of the inflationary expansion, while the correlation in
the model with initial $M^2_{\pi} = 0$ does. Therefore, the
$j$-dependence of the mode-mode correlation is practical to test
the dynamical predictions about the evolution ($t_e$-dependence)
of the primordial perturbations. Since the resolution of DWT
basically can be refined as required by the problem in hand, one
is capable of exploring the physics of the very early universe by
means of the mode-mode correlations in phase space.

%fig8

\section{Conclusion}

Using the $O(N)$ model as an example, we have investigated the
self-interacting effects on the primordial perturbations using a
self-consistent off-equilibrium approach. It is known that most
information of the dynamics and initial conditions of the system
will be lost during the non-equilibrium evolution. The
non-equilibrium evolution can only be probed via the observable
remains produced from such a process. In this paper, we have
shown that the off-equilibrium evolution of cosmological inflation
may have two such observable ``remains''.

The first one is the running spectral index of primordial perturbations 
induced by the scale-dependence due to the self-interaction of the 
inflaton.  We found that the running spectral index depends essentially 
on the rate of the particle creation and the energy transfer from the 
background to the inflaton fluctuations as well as the evolution of the 
background field. It is a signature of the energy transferring dynamics 
during the inflation. Although the running index of the $O(N)$ model is 
small comparing to current data, the negative running (Fig. 4) does 
coincide with the WMAP observation.

The second remain is the correlation function between phase space
modes of the density perturbation. Under the influence of the
self-interaction, fluctuations created from the background field
are no longer white noises. Although the one-point distribution
function of fluctuations in each Hartree mode is Gaussian, the
power of the fluctuations is scale-dependent, which gives rise to
the correlation between the phase space modes of the energy
density perturbation. Moreover, since the dynamical evolution of
the correlation depends upon the initial conditions of the
inflation, the mode-mode correlation of density perturbations also
provides a window to study the dynamics of the self-interaction
{\it as well as} the initial conditions of the inflation. Here, we
would like to emphasis two important results. First, the
initial-condition-dependency of correlation functions is
irrelevant to the number of fields in action. Second, the
inflationary density perturbations after the superhorizon
evolution are not the fluctuations of thermal equilibrium state.
Therefore, the dependence of the correlation function upon the
initial conditions does not contradict with the reheating of the
universe, which is generally produced by the interactions between
$\phi$ and other fields. Thus, we may expect that the non-trivial
mode-mode correlation in the phase space is detectable via a DWT
analysis on the CMB temperature map, or other observations on the
large scale structure relevant to the density
perturbations~\cite{pvf}.

Although, the $O(N)$ model is just an example to illuminate the
off-equilibrium effects from the self-interaction of the inflaton
field, we believe the implication from what we have found in this paper
is also useful to other interacting inflation models. For instance,
the scale-dependence of perturbations drawn from the self-consistent
off-equilibrium dynamics is generally different from that obtained by
the third or the higher derivative of a classical effective potential.
Therefore, the running spectral index determined by the formalism of
various effective potentials would be questionable if the interaction or
self-interaction of the inflaton is substantial.

\acknowledgments

This work was supported in part by the National Science Council,
Taiwan, ROC under the Grant NSC92-2112-M-001-029 (WLL), NSC92-2112-M-001-030
(YYC) and NSC91-2112-M-259-005 (DSL), and by the TWQCG Quantum Vacuum
Workshop of the National Center of Theoretical Sciences
(NSC 91-2119-M-007-004-),
Taiwan, R.O.C. We thank Long-Long Feng for providing many helps in the
DWT calculations. DSL would like thanks Hector de Vega and Bei-Lok Hu for
their stimulating discussions.

%\appendix


\begin{thebibliography}{99}

\bibitem{lid} A. R. Liddle and D. H. Lyth, Cosmological Inflation
 and Large Scale Structure, (Cambridge, 2000).

\bibitem{pei} H. V. Peiris, {\em et al.},''First Year Wilkinson
Microwave Anisotropy Probe (WMAP) Observations: Implications for
Inflation'',  astro-ph/0302225.

\bibitem{kaw} M. Kawasaki, M. Yamaguchi, and  J. Yokoyama, Phys.Rev. D
{\bf 68}, 023508 (20030; J. H. Chung, G. Shiu, and M. Trodden,
Phys. Rev. D {\bf 68}, 063501 (2003) ; J. M. Lidsey and R.
Tavakol, ''Running of the Scalar Spectral Index and Observational
Signatures of Inflation'', astro-ph/0304113;  Bo Feng, Mingzhe Li,
Ren-Jie Zhang, and Xinmin Zhang, ''An inflation model with large
variations in spectral index'', astro-ph/0302479.

\bibitem{do} L. Doland and R. Jackiw, Phys. Rev. D {\bf 9} 3320
 (1974).

\bibitem{we} E. J. Weinberger and A. Wu, Phys. Rev. D {\bf 36}, 2747
 (1987).

\bibitem{huang} Z. Huang and X. N. Wang, Phys. Rev. D. {\bf 49}, 4335 (1994).

\bibitem{sam} O. \'Eboli, R. Jackiw, and S.-Y. Pi, Phys. Rev.
  {\bf D37}, 3557 (1988); M. Samiullah, O. \'Eboli, and  S.-Y. Pi, Phys. Rev.
  {\bf D44}, 2335 (1991).

\bibitem{baa}  J. Baacke, K. Heitmann, and C. Patzold, Phys. Rev.
   {\bf D55}, 2320 (1997); J. Baacke and K. Heitmann, Phys. Rev.
   {\bf D62}, 105022 (2000); J. Baacke and S. Michalski, Phys. Rev.
   {\bf D65}, 065019 (2002); J. Baacke and A. Heinen, hep-ph/0311282.

\bibitem{boyn1} D. Boyanovsky,  D.-S Lee, and  A. Singh, Phys. Rev.
   {\bf D48}, 800 (1993); D. Boyanovsky, H. J. de Vega, R. Holman,
    D.-S Lee, and A. Singh, Phys. Rev.
   {\bf D51}, 4419  (1995); D. Boyanovsky, M. D'Attanasio, H. J. de Vega,
    R. Holman, and  D.S. Lee, Phys. Rev. {\bf D52}, 6805 (1995); D.
    Boyanovsky, I. D. Lawrie and D.-S. Lee, Phys. Rev. {\bf D54}, 6013
    (1996).

\bibitem{boyn2}D. Boyanovsky, D. Cormier, H. J. de Vega, and  R. Holman,
 Phys. Rev. {\bf D55}, 3373 (1997); D. Boyanovsky, D. Cormier, H.
J. deVega, R. Holman, and S. P. Kumar, Phys. Rev. {\bf D57}, 2166
(1998);

\bibitem{ra} S. A. Ramsey and B. L. Hu, Phys. Rev. {\bf D56}, 661 (1997).

\bibitem{sta} D. Polarski and A. A. Starobinsky, {\it Class. Quantum. Grav.}
  {\bf 13}, 377 (1996); J. Lesgourgues, D. Polarski, and A. A. Starobinsky,
{\it Nucl. Phys.} {\bf B497}, 479 (1997).

\bibitem{lee} W. L. Lee and L. Z. Fang, Europhys. Lett. {\bf 56},
   904 (2001).

\bibitem{re} S.-J. Rey, Nucl. Phys. {\bf B284}, 706 (1987); R. Brandenberger,
   R. Laflamme, and M. Mijic, Mod. Phys. {\bf A 5}, 2311 (1990).

\bibitem{coo} F. Cooper, S. Habib, Y. Kluger, and  E. Mottola, Phys. Rev.
   {\bf D55}, 6471 (1997); D. L. Kaiser, Phys. Rev. {\bf D59}, 117901 (1999);
   H. Hiro-Oka and H. Minakata, Phys. Rev. {\bf C64}, 044902
   (2001).

\bibitem{dau} Y. Meyer, {\it Wavelets and Operators}, (Cambridge, 1992);
 I. Daubechies, {\it Ten Lectures on Wavelets}, (SIAM, 1992); L. Z. Fang
\& R. Thews, {\it Wavelets in Physics}, (World Scientific, 1998).

\bibitem{ko} E. W. Kolb and M. S. Turner,{\it The Early Universe}
 (Addison-Wesley, Reading, MA,1990).

\bibitem{pvf} J. Pando, D. Valls--Gabaud and L. Z. Fang, Phys. Rev. Lett.,
     {\bf 81}, 4568 (1998).

\end{thebibliography}
\end{document}